\documentclass[aps,onecolumn,11pt,floatfix,altaffilletter,tightenlines,showpacs,showkeys,notitlepage,nofootinbib]{revtex4-1}

\usepackage[colorlinks=true,citecolor=blue,linkcolor=blue]{hyperref}
\usepackage[normalem]{ulem}
\usepackage{amsmath,amssymb}
\usepackage{epsfig}
\usepackage{graphicx}               
\usepackage{url}
\usepackage{color}
\usepackage{multirow}
\usepackage{placeins}
\usepackage[dvipsnames]{xcolor}
\usepackage{epstopdf}
\usepackage{dsfont}
\usepackage{tikz}
\usepackage{bbold}
\usepackage{cases}
\usepackage{empheq}
\usetikzlibrary{trees}
\usetikzlibrary{decorations.pathmorphing}
\usetikzlibrary{decorations.markings}
\usepackage{soul}

\definecolor{jblue}  {RGB}{20,50,100}
\definecolor{npurple}  {RGB} {153, 51, 204}
\definecolor{wred}   {RGB}{217,0,56}
\definecolor{white}   {RGB}{255,255,255}

\definecolor{korange}   {RGB}{235, 80,  43}
\definecolor{korange2}   {RGB}{245, 100,  63}
\definecolor{kyelloworange}   {RGB}{255, 210,  110}
\definecolor{kyelloworange2}   {RGB}{240, 170,  90}
\definecolor{kred}   {RGB}{204,  102, 153}
\definecolor{kpurple}   {RGB}{153,  61, 190}
\definecolor{kpurplelight}   {RGB}{213,  161, 230}

\usepackage[capitalize]{cleveref}
\usepackage{subfigure} 
\usepackage{lipsum}
\definecolor{red}{rgb}{1.0, 0, 0}
 \usepackage{gensymb}
 
\allowdisplaybreaks

\bibpunct{[}{]}{,}{n}{}{,}

\renewcommand{\vec}[1]{{\mathbf{#1}}}


\begin{document}

\title{Low Scale Left-Right Symmetry and Naturally Small Neutrino Mass
 }

\author{Vedran Brdar}   \email{vbrdar@mpi-hd.mpg.de}
\author{Alexei Yu. Smirnov} \email{smirnov@mpi-hd.mpg.de}
\affiliation{Max-Planck-Institut f\"ur Kernphysik,
       69117~Heidelberg, Germany}

\begin{abstract}
\noindent
We consider the low scale ($10$ - $100$ TeV) left-right symmetric model
with ``naturally'' small neutrino masses generated through the inverse seesaw mechanism. 
The Dirac neutrino mass terms are taken to be similar to the masses of charged 
leptons and quarks in order to satisfy the quark-lepton similarity condition. 
The inverse seesaw implies the existence of fermion singlets 
$S$ with Majorana mass terms as well as the ``left" and ``right" Higgs doublets.
These doublets provide the portal for $S$ and break the left-right symmetry.
The inverse seesaw allows to realize a scenario in which
the large lepton mixing originates from the Majorana mass matrix of
$S$ fields which has certain symmetry.
The model contains
heavy pseudo-Dirac fermions, formed by $S$ and the right-handed neutrinos, 
which have masses in the $1$ GeV - $100$ TeV range 
and can be searched for at current and various future colliders such as LHC, 
FCC-ee and FCC-hh as well as in SHiP and DUNE experiments.
Their contribution to neutrinoless double beta decay is unobservable. 
The radiative corrections to the mass of the Higgs boson and 
the possibility for generating the baryon asymmetry of the Universe 
are discussed. Modification of the model with two singlets ($S_L$ and $S_R$)
per generation can provide a viable keV-scale dark matter candidate.

\end{abstract}
\maketitle

\section{Introduction}
\label{sec:intro}
\noindent
The left-right symmetric models \cite{Pati/Salam,Mohapatra/Pati,Mohapatra/Pati2,Goran/Mohapatra,MS} 
based on the gauge symmetry group 
\begin{align}
SU(2)_L\times SU(2)_R \times U(1)_{B-L}\,,
\label{eq:symmetry}
\end{align} 
and parity $P$ which ensures the equality between couplings in the left and right sectors,   
are still one of the most appealing and well-motivated extensions 
of the Standard Model (SM) \cite{Senjanovic:2016bya,Dev:2014xea,Lindner:2016lxq,Dev:2016dja,Dev:2009aw,Tello:2010am,Zhang:2007da,
Das:2016akd,Das:2017hmg,FileviezPerez:2016erl}.
Spontaneous breaking of $SU(2)_R \times U(1)_{B-L} \times P$ down to the 
SM symmetry group explains the  observed low-energy asymmetry 
between the left and right as well as  provides a natural framework 
for the generation of small neutrino masses via the seesaw 
mechanism~\cite{Minkowski,GellMann:1980vs,Yanagida:1979as,Goran}.
The scale  of left-right (L-R) symmetry breaking and the seesaw scale coincide.
The key question is whether this scale can be  
at $\mathcal{O}(10 -100)$ TeV energies that are 
accessible to  LHC and the next generation of colliders. 

The majority of the low scale L-R symmetric models constructed so far is at odds with the generation 
of ``naturally" small neutrino masses. In what follows we will call the neutrino mass to be naturally small if mechanism of its generation employs
the Dirac neutrino masses, $m_\nu^D$, similar in size to the Dirac masses 
of charged leptons, $m_l$,  and quarks, $m_q$, {\it{i.e.}}  
\begin{equation}
m_\nu^D \approx m_q, m_l\,.
\label{eq:normal}
\end{equation}
This relation facilitates the grand unification and we will refer to it as to the quark-lepton ($q$-$l$) similarity condition. 
The usual type-I seesaw mechanism realizes such a possibility provided that the scale of right-handed 
(RH) neutrino masses is about $10^{14}$ GeV. Lowering the scale of RH neutrinos down to {\it{e.g.}} 
$10^4$ GeV  requires $m_\nu^D  \sim  1$ MeV, which is $5$ orders 
of magnitude smaller than the top quark mass and thus is not in accord with \cref{eq:normal} 
for the third generation of neutrinos. If neutrinos acquire their masses 
dominantly via the type-II seesaw mechanism, the Dirac mass terms  should 
be even more strongly suppressed.

In this paper, in order to reconcile the
low scale L-R symmetry and the naturally small neutrino masses 
we assume that the latter are generated via the inverse seesaw 
mechanism \cite{inverse,inverse2}.  
In such a framework, the small
neutrino masses can be  obtained for values of the Dirac mass terms that are in accord with \cref{eq:normal}.
The inverse seesaw mechanism requires the introduction of new fermionic singlets, 
$S$, which couple to the RH neutrinos and thus form the Dirac mass terms.
We introduce three such singlets (one per generation), 
whose Majorana masses are much smaller than the electroweak scale. 
The generation of light neutrino masses 
via the inverse seesaw requires the right-handed $SU(2)_{R}$ Higgs doublet and 
therefore, due to L-R symmetry, the left-handed $SU(2)_{L}$ doublet. 
These doublets break the L-R symmetry, so that 
the Higgs triplets are not needed \cite{Gu:2010xc,Akhmedov:1995ip,Wyler:1982dd}.
Hence, the problem of the absence of low-dimensional representations\footnote{
In the vast majority of studies, L-R symmetry is broken by introducing 
the Higgs triplets, while the Higgs doublets are absent.
The existence of the higher representations
(triplets) and the absence of low-dimensional representations
(doublets here) should have a certain reason and a proper physical explanation.} in the model does not appear. 

We do not assume any special smallness of
the Yukawa couplings of the scalar doublets. These couplings are similar or even equal to the couplings 
of the bi-doublet. If equal, the so called screening of the Dirac structures is realized~\cite{Lindner:2005pk},  
and the large lepton mixing originates from  a certain structure  of the Majorana 
mass matrix of singlets $S$ (the screening was previously studied for the double seesaw mechanism \cite{Smirnov:2018luj}). The mass matrix of $S$  may have certain symmetry which leads, {\it{e.g.}}, to the tribimaximal mixing.

In this paper we elaborate on such a scenario.
We focus on generation of neutrino mass and mixing in
the L-R model with Higgs doublets instead of triplets. Such models have
been extensively explored before \cite{Goran/Mohapatra,Senjanovic:1978ev} and the only new element here are 
singlet fermions $S$ which allow to realize the inverse seesaw mechanism (as a dominant mechanism for the generation of neutrino mass) and certain selection of the Yukawa couplings.
 We explore here new features related to introduction of the fermion singlets, while the gauge and scalar boson sectors are the same as in several earlier publications. We confront the model with the existing 
experimental data from the beam-dump experiments, LHC, experiments searching for neutrinoless double beta decay, {\it  etc}.,  
and also estimate the discovery potential of future colliders and neutrino oscillation facilities. 
We obtain bounds on relevant parameters and constrain the L-R symmetry breaking scale.  
We examine the possibility for the generation of the observed baryon asymmetry 
of the Universe and address the issue of the Higgs naturalness.
We also consider the extensions of the aforementioned scenario, in particular the scenario with two $S$ fields (left and right) per generation which contains a viable dark matter candidate.

The paper is organized as follows. 
In \cref{sec:model}, we describe the model and generation of the neutrino masses, and discuss the possibility to introduce flavor symmetries. 
The phenomenology of the model is presented in \cref{sec:pheno}. 
We elaborate on the various extensions of this scenario in \cref{sec:beyond}.
Finally, in \cref{sec:summary} we conclude.

\section{The Model and neutrino masses}
\label{sec:model}
\subsection{The model, linear and inverse seesaw}
\noindent
Leptons are organized in the following representations of the symmetry group (\ref{eq:symmetry})
\begin{equation}
 L_L =  \begin{pmatrix}\,\nu_l\, \\[0.07cm] \,l\, \end{pmatrix}_L \sim (2,1,-1)\,,\qquad
 L_R =  \begin{pmatrix}\,\nu_l\,\\[0.07cm] \,l\, \end{pmatrix}_R \sim (1,2,-1)\,, \qquad
  S \sim (1,1,0)\,,
  \label{eq:lepton_doublets}
\end{equation}
where $l = \{e, \mu, \tau\}$ and in brackets we indicate the corresponding quantum numbers. The Majorana leptons $S$ are  complete gauge singlets.
We assume the existence of three such leptons - one per generation. 

The scalar sector consists of the usual bi-doublet  
\begin{align}
\Phi = \begin{pmatrix}\phi_1^0 & \phi_2^+ \\[0.1cm]
\phi_1^- & \phi_2^0\end{pmatrix}
\sim \left(2,2,0\right),
\label{eq:bidoublet}
\end{align}
and two doublets 
\begin{align}
~~~~~~~~~~~~~~~~~~~~
&\chi_L = \begin{pmatrix}\chi_L^+ \\[0.1cm] \chi_L^0 \end{pmatrix}\sim (2,1,1)\,, &
\chi_R = \begin{pmatrix}\chi_R^+ \\[0.1cm] \chi_R^0 \end{pmatrix}\sim(1,2,1)\,.
~~~~~~~~~~~~~~~~~~~~
\label{eq:scalar_doublets}
\end{align}
The latter are required for the realization of the inverse seesaw mechanism as well as for  the L-R symmetry breaking.   
 In the first step, the neutral component of the right-handed scalar 
doublet $\chi_R$ obtains a non-vanishing vacuum expectation value (VEV), which breaks the 
$SU(2)_R\times U(1)_{B-L}$ down to $U(1)_Y$, where $Y$ is the  hypercharge. In the second step, the
neutral scalar fields from left-handed scalar doublet $\chi_L$ ($\chi_L^0$) 
and  bi-doublet $\Phi$ ($\phi_1^0,\,\phi_2^0$) break 
the $SU(2)_L \times U(1)_Y$ symmetry down to $U(1)_{\text{EM}}$. 
Their VEVs should satisfy the  relation 
\begin{align}
 \sqrt{ \langle \chi_L^0 \rangle^2 + \langle \phi_{1}^0 \rangle^2 
+ \langle \phi_{2}^0 \rangle^2} \approx 246\,\text{GeV},
\end{align}
in order to reproduce the electroweak scale.
Notice that  for the electroweak symmetry breaking 
only one non-zero VEV among these three neutral fields is enough. The L-R symmetry breaking requires
\begin{align}
\langle \chi_R^0 \rangle\gg \langle \chi_L^0 \rangle, \langle \phi_{1,2}^0 \rangle\,. 
\label{eq:vevs}
\end{align}


The Higgs sector in \cref{eq:bidoublet,eq:scalar_doublets} is identical to the one in earlier publications \cite{Goran/Mohapatra,Senjanovic:1978ev}. Minimization of the potential had been done under certain
simplifications. In particular, it was assumed {\it a priori}  that the
electic charge is conserved in minimum. The tri-linear terms in potential are absent due to additional symmetry. With such conditions, it was shown \cite{Goran/Mohapatra,Senjanovic:1978ev}
that the global minimum exists for a certain range
of parameters of the potential, in which the inequality (\ref{eq:vevs}) is satisfied, {\it{i.e.}} the parity is spontaneosly broken. The requirements for such a scenario are
inequalities of certain quartic couplings and positivity or small values of other couplings \cite{Goran/Mohapatra,Senjanovic:1978ev}. The values of VEVs are controlled (at least in the case of mild hierarchy)  by the
mass parameters for the doublets and bi-doublet.
For the most general form of the potential (and without above discussed assumptions) the minimization has not been done for the bi-doublet--doublet case. Such a study has been done recently \cite{Dev:2018foq} for
the bi-doublet--triplet scenario and some information can be inferred from the obtained results. It has been found in \cite{Dev:2018foq} that there are significant regions in the parameter space where required minimum can be obtained. Note that bi-doublet--doublet and bi-doublet--triplet
models have the same number of neutral bosons, and structure of the
terms in the potential is similar. Interestingly, the $\beta-$ terms
in the bi-doublet--triplet potential should be small in the
minimum and the terms of similar type in the bidoublet--doublet potential are absent.

For the rest of our  paper it is enough that
hierarchy (\ref{eq:vevs}) is achieved at least for some choice of parameters.
Since we are not discussing phenomenology of the Higgs sector
(masses of scalars, decay rates, etc.)
specific values of the parameters of the higgs potential are not
important.

The lepton masses are generated by the following Lagrangian  \cite{Gu:2010xc}
\begin{align}
\mathcal{L}\supset - \bar{L}_R \,Y\, \Phi^\dagger L_L - 
\bar{L}_R \,\tilde{Y}\, \tilde\Phi^\dagger L_L  -
\bar{S} \, Y_L  \tilde{\chi}_L^\dagger L_L-
 \bar{S}^c \, Y_R \, \tilde{\chi}_R^\dagger L_R  -\frac{1}{2}  
\,\bar{S}^c \, {\mu}\, S + \text{h.c.}\,,
\label{eq:lag_fermion}
\end{align}    
where  $Y$, $\tilde Y$, $Y_L$, $Y_R$ are $3\times 3$ matrices of the Yukawa couplings 
and $\mu$ is the $3\times 3$ Majorana mass matrix of $S$ leptons.
$\tilde{X} \equiv i\sigma_2 X^*$ $\left(X=\{\chi_L,\chi_R\}\right)$, $S^c \equiv C \bar{S}^T$ and $\tilde{\Phi} \equiv \sigma_2 \Phi^* \sigma_2$ denote charge conjugated fields of scalars and fermions. The field transformations under parity
\begin{align}
~~~~~~~~~~~~~~~~~~~L_L &\Longleftrightarrow L_R \,,& \chi_L &\Longleftrightarrow \chi_R \,,& \Phi &\Longleftrightarrow \Phi^\dagger\,,& S &\Longleftrightarrow S^c\,,~~~~~~~~~~~~~~~~~~~
\label{eq:field_tr}
\end{align}
impose the following relations 
\begin{align}
~~~~~~~~~~~~~~~~~~Y_L &= Y_R\,, & Y&=Y^\dagger\,, & \tilde{Y}&=\tilde{Y}^\dagger\,, & \mu&=\mu^\dagger\,,~~~~~~~~~~~~~~~
\label{eq:parity_rel}
\end{align}
above the L-R symmetry breaking scale.

When the scalar fields acquire VEVs, the interactions (\ref{eq:lag_fermion}) generate the  mass matrix of neutral leptons 
\begin{align}
{\cal M} = 
\begin{pmatrix}
0  & m_D^T  & {m_D'^T}\,  \\[0.1cm]
 m_D  & 0  & M_D^T  \\[0.1cm]
m_D'  & M_D & {\mu} 
\end{pmatrix}\,,
\label{eq:neutral_mass_1}
\end{align}
given in the $(\nu_L, N_L, S^c)$ basis $(N_L\equiv\nu_R^c)$. Here, 
\begin{align}
m_D =\frac{1}{\sqrt{2}}\left(Y \langle \phi_1^0 \rangle\, + 
\tilde Y \langle \phi_2^0  \rangle \,\right), \qquad
m_D' =  \frac{1}{\sqrt{2}} Y_L\langle  \chi_L^0 \rangle \, ,   \qquad
M_D = \frac{1}{\sqrt{2}} Y_R \langle \chi_R^0 \rangle\,.
\label{eq:r}
\end{align}
For simplicity  we assume 
 $\langle \phi_1^0 \rangle \gg  \langle \phi_2^0 \rangle$ such 
that only the first term in $m_D$  contributes.

The block diagonalization of ${\cal M}$ leads to the light 
neutrino mass matrix 
\begin{align}
m_\nu \simeq \frac{{\langle \chi_L^0 \rangle}}{\langle \chi_R^0 \rangle} 
\left(m_D + m_D^T\right) -  \, m_D^T\, 
M_D^{-1}\, {\mu}\,\left(M_D^T\right)^{-1}\, m_D\,.
\label{eq:neutrino_mass_inv+lin}
\end{align}
Here, the first term is the linear seesaw contribution \cite{inverse},  
whose existence is a generic consequence of the inverse seesaw realization in the L-R models. Notice that due to $Y_L= Y_R$, the Yukawa matrices cancel and this term is given by the Dirac mass matrix 
multiplied by the ratio of VEVs of the two doublets. 
For ${\langle \chi_L^0 \rangle}$ of the order of electroweak scale, 
and in the absence of unnaturally small elements of $m_D$, this term yields too large neutrino masses. Furthermore, if $m_D\propto m_u$ (the subscript $u$ denotes the up-type quarks),
it has a wrong flavor structure with too strong mass hierarchy and small mixing. 
This is incompatible with the neutrino mass squared differences and large mixing angles 
observed in the oscillation experiments. Therefore, the linear seesaw contribution 
should be at most sub-dominant and the main contribution to the neutrino mass should arise 
from the inverse seesaw, given in the second term of \cref{eq:neutrino_mass_inv+lin}. 
Since $m_D \sim m_u$, this is achieved for 
\begin{align}
\frac{\langle \chi_L^0 \rangle}{\langle \chi_R^0 \rangle} < 
\frac{0.05\,\text{eV}}{2\, m_D^{max}} \sim  10^{-12},
\label{eq:ratio}
\end{align}
where $m_D^{max}$ denotes the largest entry of the Dirac neutrino mass matrix.
It is worth noting that for $\langle \chi_L^0\rangle, \,\langle \phi_2^0 \rangle
\ll  \langle \phi_1^0 \rangle \approx 246$\,GeV, the SM Higgs boson is associated to the real part of $\phi_1^0$ field.

In order to estimate $\langle \chi_L^0 \rangle$ we consider the following terms of the scalar potential
\begin{align}
V\supset h \chi_L^\dagger \tilde{\Phi} \chi_R - m_{\chi}^2 \chi_L^\dag \chi_L\,,
\label{eq:toy_potential}
\end{align}
where $h$ is the dimensionful coupling and   
the term $\lambda (\chi_L^\dagger \chi_L)^2$ can be neglected for small values of  
 $\langle\chi_L^0\rangle$.
The minimization condition, $\partial V/ \partial \chi_L^\dag=0$, gives 
\begin{align}
\langle \chi_L^0 \rangle = h\,\frac{\langle \phi_1^0 \rangle}{\langle \chi_R^0 \rangle}\,,
\label{eq:vev-seesaw}
\end{align}
where we have taken into account that  due to the L-R symmetry
$m_{\chi_L}^2 =  m_{\chi_R}^2 \sim \langle \chi_R^0 \rangle^2$. 
Using the condition (\ref{eq:ratio}) we obtain 
\begin{align}
h\lesssim 40 \,\text{keV} \left(\frac{\langle \chi_R^0 \rangle}{10^5~ \text{GeV}} \right)^2,
\label{eq:h}
\end{align}
which needs to be satisfied in order to generate neutrino masses mainly via the inverse seesaw mechanism. Notice that \cref{eq:vev-seesaw} is a realization of the VEV seesaw \cite{Deshpande:1990ip}.
In contrast to the triplet case, the electroweak scale VEV (of the bi-doublet)
enters this relation linearly and it contains dimensionful coupling $h$.
According to \cref{eq:vev-seesaw}, the VEV of $\chi_L^0$ is controlled by
free parameter $h$. 

\begin{figure}
  \centering
  \begin{tabular}{ccc}
    \includegraphics[width=0.45\textwidth]{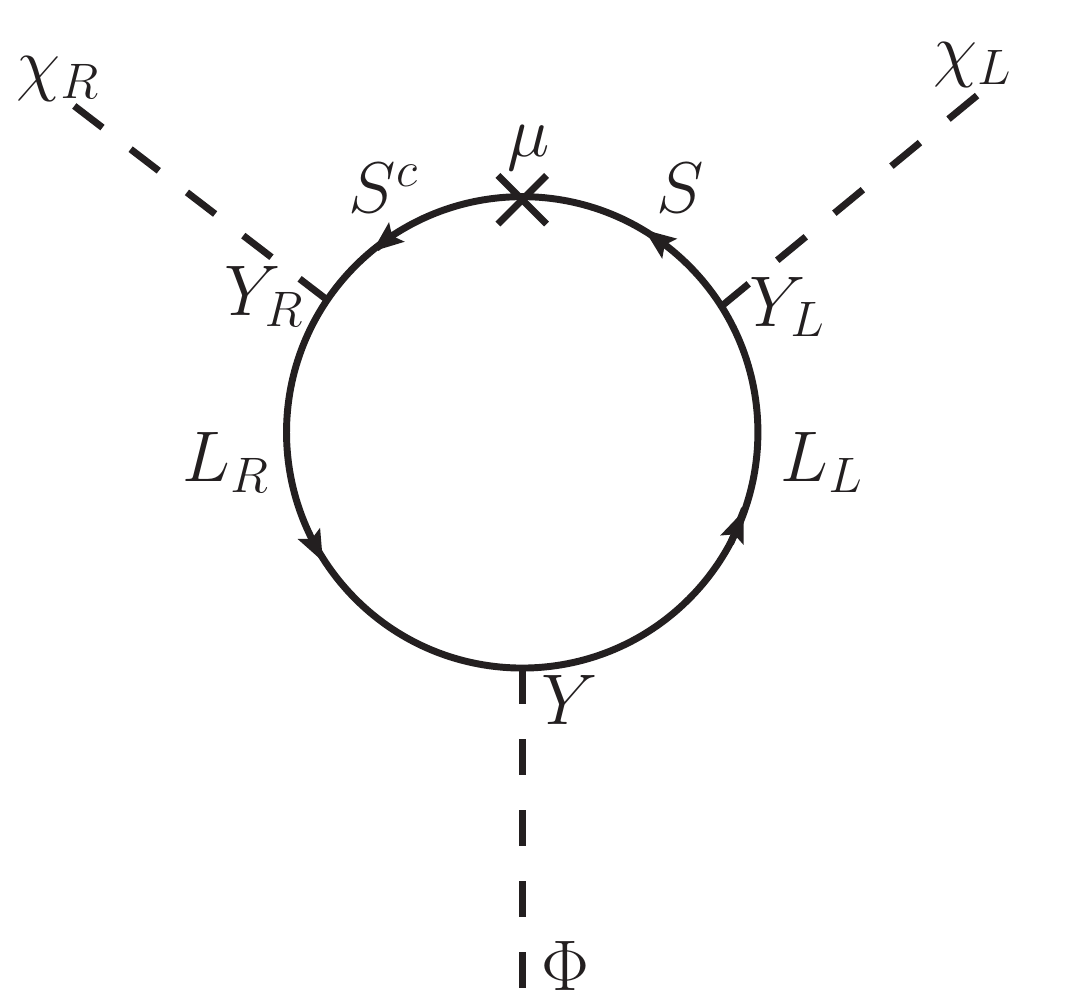}
  \end{tabular}
  \caption{1-loop diagram generating $\chi_L^\dagger \Phi \chi_R$ term in the potential.
}
  \label{fig:vev}
\end{figure}

The coupling $h$ in the potential (15) can be forbidden by symmetry
with respect to transformation $\Phi \rightarrow e^{i\pi/2} \Phi$
\cite{Senjanovic:1978ev}.
This symmetry is explicitly broken by the Yukawa interactions of $\Phi$,
and therefore it does not prevent the appearance of the $h$ term in
higher orders of perturbation theory.
Even if the coupling $h$ vanishes at tree-level due to some symmetry, 
$\langle \chi_L^0 \rangle \neq0$ is generated radiatively via the one-loop diagram shown in \cref{fig:vev}. It can be estimated as 
\begin{align}
\langle \chi_L^0 \rangle \simeq  \frac{1}{16 \pi^2}  
\langle \chi_R^0 \rangle \frac{\langle \phi_1^0 \rangle}{\langle \chi_R^0 \rangle} \frac{\mu}{\langle \chi_R^0 \rangle}. 
\label{eq:vev_L}
\end{align}
For $\mu \simeq \mathcal{O}(10-100)$ keV (see below) and $\langle \chi_R^0 \rangle = 10^5$ GeV \cref{eq:vev_L} gives 
$\langle \chi_L^0 \rangle \simeq 10^{-14} \langle \chi_R^0 \rangle$ 
which satisfies \cref{eq:ratio}. 

Notice that for $\langle \chi_R^0 \rangle \sim 10^5$ GeV (see \cref{subsec:sterile}),
$h$ should be at most 100 keV, which (as we will see) is  of the order of $\mu$. 
This smallness can be associated to  violation of the lepton number. 
In particular, one can reintroduce the global lepton number $L_g$ and 
assign the charges $\left(1, -1, 1\right)$ to  $\left(\nu_L, N_L, S\right)$.  
In the limit $m_D', \mu \rightarrow 0$, the conservation of  $\tilde{L}$ is recovered 
and hence the small value of $\mu$ appears to be technically natural \`a la 't Hooft \cite{tHooft:1979rat}.

The inverse seesaw contribution  
in \cref{eq:neutrino_mass_inv+lin}  can be rewritten as  
\begin{align}
m_\nu \approx \frac{\langle \phi_1^0 \rangle^2}{\langle \chi_R^0 \rangle^2} \, 
Y^T\, Y_R^{-1}\, {\mu}\,\left(Y_R^T\right)^{-1}\, Y\,.
\label{eq:nu_mass_inv}
\end{align}
For $Y_3, {Y_{R}}_3 = 1$ we obtain $m_\nu \approx (\langle \phi_1^0 \rangle/\langle \chi_R^0\rangle)^2 \,\mu$ 
(hereafter we denote Yukawa matrices and the corresponding Dirac mass 
terms with only one subscript since these matrices are taken to be diagonal (see \cref{sec:flavor})) 
which allows us to estimate $\langle \chi_R^0 \rangle$ as
\begin{align}
\langle \chi_R^0 \rangle \simeq  3.5 \cdot 10^{5}~ {\rm GeV} \left(\frac{\mu}{100 \text{keV}}\right)^{1/2}\,
\left(\frac{0.05\,\text{eV}}{m_{\nu 3}} \right)^{1/2}\,,
\label{eq:massrh}
\end{align}
where $\mu$ is the representative element of the corresponding matrix and $m_{\nu3}$ is  the largest active neutrino mass.
Consequently, 
\begin{align}
{\mu}_{ij} \ll {m_D}_{i}\,,{M_D}_{i}\,. 
\end{align}
 
The inverse seesaw  not only explains the smallness of neutrino masses 
under the condition of $q$-$l$ similarity, but also provides a rather appealing framework for the lepton mixing generation where the large mixing angles originate from
the $\mu$ matrix of singlets $S$, {\it{i.e.}} from the hidden sector. 

The leptonic mixing matrix (PMNS) can be written as
\begin{align}
U_{\text{PMNS}} = U_l^{\dagger} U_\nu\,, 
\label{eq:pmns-rep}
\end{align}
where $U_{l}$ follows from the diagonalization
of the charged lepton mass matrix, whereas
$U_\nu$ diagonalizes the neutrino mass matrix (\ref{eq:nu_mass_inv})
generated by the inverse seesaw mechanism.
A good agreement with experimental data can be achieved if
\begin{align}
~~~~~~~~~~~~~~~~~~U_{l} &\approx V_{\text{CKM}}\,,& 
U_\nu &\sim U_{\text{TBM}}\,\,\text{or}\,\, U_{\text{BM}}\,,~~~~~~~~~~~~~~~
\label{eq:matrices}
\end{align}
where $V_{\text{CKM}}$ is the mixing matrix in the quark sector, and 
TBM and BM denote tribimaximal \cite{Harrison:2002er} and bimaximal \cite{Vissani:1997pa,Barger:1998ta} mixing matrices.
The first relation in \cref{eq:matrices} could be a consequence 
of the grand unification or identical horizontal symmetry 
in the quark and lepton sectors. The second relation can follow from  certain 
symmetry in the singlet sector.

\subsection{Screening and $q$-$l$ similarity}
\label{subsec:q-l}
\noindent
Let us consider the generation of neutrino mixing in the basis
where  $Y$,  and therefore the Dirac mass matrix $m_D$, are diagonal
\begin{equation}
Y = Y^{\text{diag}} \equiv \text{diag}(Y_1,~ Y_2,~ Y_3 )\,.
\label{eq:ymatrix}
\end{equation}

First, we assume that
\begin{equation}
Y_R =  Y   
\label{eq:scren}
\end{equation}
(and also $Y_L  = Y$),
so that $m_D \propto M_D$. 
This equality can be a consequence (a remnant) of further unification 
when $\nu_L$, $N_L$ and $S^c$ enter the same multiplet,
{\it{e.g.}} 27-plet of the $E_6$ grand unification theory. This relation can also stem from certain horizontal symmetry \cite{Ludl:2015tha}. The equality in \cref{eq:scren} leads to the screening
of the Dirac structures: $Y_R$ and $Y$ cancel in the expression for the light neutrino mass matrix (see \cref{eq:nu_mass_inv}), so that
\begin{align}
m_\nu \approx \xi^2 {\mu}\,, 
\label{eq:lightmm}
\end{align}
where 
\begin{align}
\xi \equiv  \frac{\langle \phi_1^0 \rangle}{\langle \chi_R^0 \rangle} = \frac{{m_D}_{i}}{{M_D}_{i}}\,.
\end{align}
According to \cref{eq:lightmm}, the structure of the light neutrino mass matrix is given 
by the structure of the Majorana matrix $\mu$.
In particular, the neutrino contribution to the PMNS matrix  
is determined  by  $\mu$:
\begin{align}
U_\nu^T\, m_\nu\, U_\nu = \xi^2\,  U_\nu^T\, {\mu} \,U_\nu = \text{diag}\left(m_{\nu 1},m_{\nu 2},m_{\nu 3}\right).
\label{eq:diag}
\end{align}
Previously, such a cancellation of the couplings has been
considered for the double seesaw model in Refs. \cite{Smirnov:1993af,Lindner:2005pk,Ludl:2015tha,Smirnov:2018luj}.

Second,  we assume the $q$-$l$ similarity 
\begin{align}
Y\approx Y_u\,,
\label{eq:equal}
\end{align}
where $Y_u$ is the up-type quark Yukawa coupling matrix. 
The screening  and the $q$-$l$ similarity conditions determine the phenomenology of this scenario.

\subsection{Flavor symmetries}
\label{sec:flavor}
\noindent

The matrices $Y$ and $Y_R$ can be diagonal simultaneously
due to the $G_{\text{basis}} =  Z_2 \times Z_2$ symmetry with $(-,-)$, $(+,-)$, $(-,+)$ charges 
for the three generations of fermions and uncharged scalar sector. We will call $G_{\text{basis}}$ the basis fixing symmetry \cite{Ludl:2015tha}.
This symmetry is broken by the non-diagonal matrix $\mu$ and, in fact, the smallness of $\mu$ 
with respect to the other scales in the model can be related to this breaking.
The $\mu$ term can arise from the interactions of $S$ with the new gauge singlet 
bosons which carry non-trivial $Z_2 \times Z_2$ charge and develop non-zero VEVs.
This Abelian symmetry, however,  does not ensure the equality of
the diagonal elements of $Y$ and $Y_R$. The equality
can be achieved by introducing, for instance,  a permutation symmetry or by further unification mentioned above.

\begin{figure}
  \centering
  \begin{tabular}{ccc}
    \includegraphics[width=0.60\textwidth]{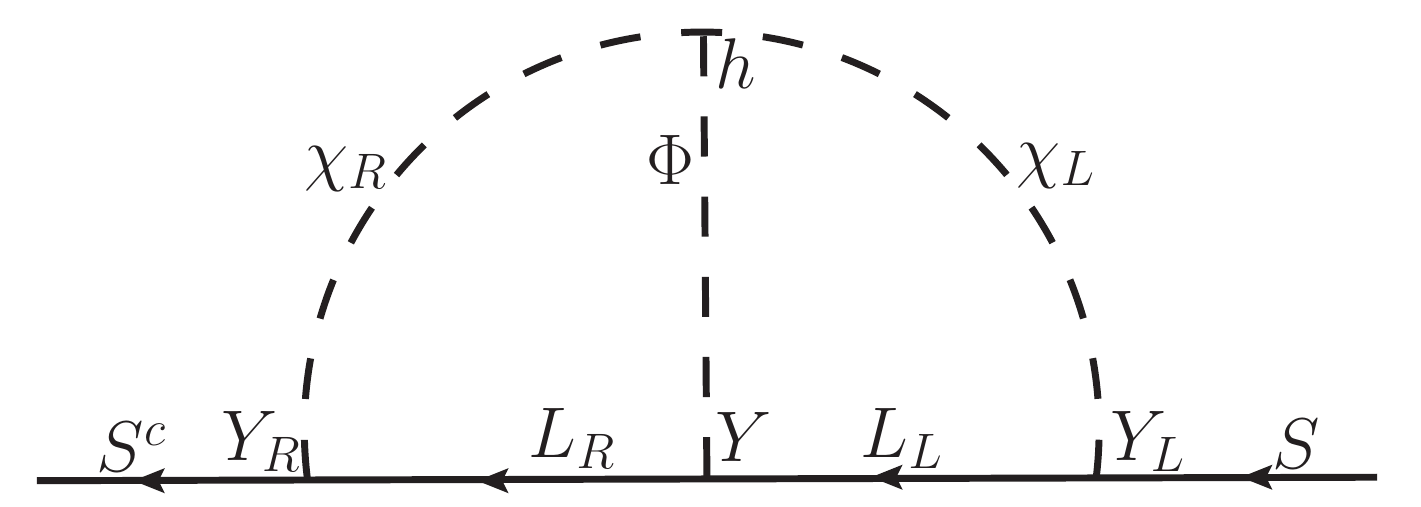}
  \end{tabular}
  \caption{The leading radiative correction to the Majorana mass $\mu$.}
  \label{fig:mu}
\end{figure}

In the case of the exact screening, the matrix $\mu$
should have nearly tribimaximal form in the basis
fixed by $G_{\text{basis}} = Z_2 \times Z_2$.
This can be achieved by introducing  the non-Abelian
({\it{e.g.}}, discrete) symmetry $G_f$ which is broken down to
$G_{\text{basis}}$ in the visible sector and to another
residual symmetry $G_{\text{hidden}} = Z_2 \times Z_2$ in the
$S-$sector \cite{Ludl:2015tha}. In the visible sector, 
$G_{f}$ can be broken explicitly. Similar construction has been
realized for the double seesaw model \cite{Smirnov:2018luj} with
masses of singlets at the Planck scale. In \cite{Smirnov:2018luj} the 
explicit symmetry breaking occurs at the lower (grand unification)
scale and its impact on the neutrino mixing is suppressed by $\mathcal{O}(M_{\text{GUT}}/M_{\text{Pl}})$ factor.
In the case of inverse seesaw, the $\mu$ scale (with certain symmetry) 
is much lower than the explicit symmetry breaking scale $M_D$. 
So, {\it a priori}, the corrections to $\mu$ can be large, thus  destroying the structure of $\mu$ imposed by symmetry.
To check this, let us assume that the required structure of $\mu$, and consequently
$m_\nu$, is achieved at the tree-level and estimate the corresponding radiative corrections. 
The lowest order correction to $\mu$  is given by the two loop diagram shown in  
\cref{fig:mu}. It  can be estimated as 
\begin{align}
\Delta \mu_{jj} \simeq  \frac{1}{(16\pi^2)^2}\, Y_{Lj}^*\, Y_{R j} \, Y_{j} \,h\,.  
\label{eq:twoloop}
\end{align}
The corrections are diagonal due to screening. The hierarchical values of Yukawa couplings 
$Y_{Lj}$,  $Y_{R j}$,  $Y_{j}$ violate symmetry that sets the pattern of $\mu$.
The largest correction is the one to $\mu_{33}$ for which $Y_{L3}= Y_{R 3} = Y_{3} =1$.
Taking $h\sim 0.1$ MeV (the value inferred from \cref{eq:h} for  
$\langle \chi_R^0 \rangle\simeq 10^5$ GeV) we obtain $\Delta \mu_{33} \sim 10$ eV.
The tree-level entries of the $\mu$ matrix are $\sim 0.1$ MeV in order to reproduce $\sim 0.1$ eV neutrino masses. 
Thus, the  radiative corrections are much smaller than the tree-level contribution 
and the structure imposed by symmetries is  preserved with high accuracy. 

\section{Phenomenology and Naturalness}
\label{sec:pheno}

\subsection{Heavy Neutral Lepton Searches}
\label{subsec:sterile}
\noindent
Neglecting the linear seesaw contribution, we obtain from \cref{eq:neutral_mass_1} 
the mass matrix in $(\nu_L, N_L, S^c)$ basis
\begin{align}
{\mathcal M} = 
\begin{pmatrix} 
0  & m_D   & 0  \\[0.05cm]
m_D & 0  & M_D  \\[0.05cm]
0   & M_D & {\mu} 
\end{pmatrix}\,, 
\label{eq:m_matrix}
\end{align}  
where both $m_D$ and $M_D$ are simultaneously diagonal under the screening assumption, 
and furthermore $m_{Di} = \xi M_{Di}$. The diagonalization of ${\mathcal M}$ can be performed in several steps. 

\noindent
$(i)$ We start with rotation in the  $\nu$-$S$  plane 
\begin{align}
{\mathcal U}_{S} = 
\begin{pmatrix}
c_\xi \, \mathbb{1}  & 0  & s_\xi \, \mathbb{1} \\[0.05cm]
 0  & \mathbb{1} & 0 \\[0.05cm]
 - s_\xi \, \mathbb{1}  & 0 & c_\xi \, \mathbb{1}
\end{pmatrix},
 \label{eq:first_rotation}
\end{align}
where 
\begin{align}
s_\xi = \frac{\xi}{\sqrt{1  +\xi^2}} \approx \xi\,,
\end{align}
and for brevity hereafter we denote $c \equiv \cos $, $s \equiv \sin $.  
After this rotation, in the new basis $(\nu',~ N_L,~ S')$, 
the  $1$-$2$ and $2$-$1$ blocks 
vanish  and the mass matrix becomes  
\begin{align}
{\cal M}_S =
\begin{pmatrix}
s_\xi^2 \,\mu  & 0   & -s_\xi \,c_\xi \,\mu  \\[0.05cm]
 0     & 0  & M_\xi  \\[0.05cm]
-s_\xi\, c_\xi\, \mu   & M_\xi  & c_\xi^2\, \mu 
\end{pmatrix},  
\end{align}
with  
\begin{align}
M_{\xi i} \equiv {M}_{Di} \sqrt{1 + \xi^2}\,.  
\end{align}
\noindent
$(ii)$ Next, we perform  rotations in the $N_L - S'$ plane  
\begin{align}
{\cal U}_{N} = 
\begin{pmatrix}
\mathbb{1} & 0 & 0 \\[0.05cm]
0 & \vec{c}_N\, \mathbb{1}  & \vec{s}_N \, \mathbb{1} \\[0.05cm]
0 & - \vec{s}_N\, \mathbb{1}  & \vec{c}_N\, \mathbb{1}
\end{pmatrix},
 \label{eq:n_rotation}
\end{align}
by the angles close to $45^\circ$ which approximately diagonalize
the $N_L$-$S'$ block. Here $\vec{s}_N \equiv \text{diag}\,(s_N^1, s_N^2, s_N^3)$ and  
\begin{equation}
s_N^i \approx \frac{1}{\sqrt{2}} \left[1 - \frac{\mu_{ii}}{4 {M_{D}}_i}\right], 
\label{eq:n_rotationang}
\end{equation}
where $\mu_{ii}\,\left(i = 1, 2, 3\right)$ are the diagonal elements of the matrix $\mu$ and we take $c_\xi\simeq 1$.  
As a result, the mass matrix in the new basis $(\nu',~ N^-,~ N^+)$ reads 
\begin{align}
{\cal M}_N \approx 
\begin{pmatrix}
s_\xi^2 {\mu} &  \frac{1}{\sqrt{2}} \,s_\xi\, {\mu}   & - \frac{1}{\sqrt{2}}\, s_\xi\,  {\mu}  \\[0.05cm]
\frac{1}{\sqrt{2}}\, s_\xi\,  {\mu}  &  M^-    & 0  \\[0.05cm]
- \frac{1}{\sqrt{2}}\, s_\xi\,  {\mu}    & 0  & {M}^+ 
\end{pmatrix}. 
\label{eq:mass}
\end{align}
Here,
\begin{align}
~~~~~~~~~M_i^- &= - {M_{\xi}}_i + \frac{1}{2} \mu_{ii}\,,& 
M_i^+ &=  {M_{\xi}}_i + \frac{1}{2} \mu_{ii}\,, ~~~~~~~~~
\label{eq:massMM}
\end{align}
are the masses of $N_i^-$ and $N_i^+$. Thus, the  fields $N_i^-$ and $N_i^+$ form  a pair of quasi-degenerate 
heavy neutral leptons  with $\sim {\mu}_{ii}$ mass splitting. 
For the phenomenology of heavy states, the $1$-$2$ and $1$-$3$ blocks in \cref{eq:mass} can be neglected.

\noindent
$(iii)$ Finally, we diagonalize the $1$-$1$ block of the matrix (\ref{eq:mass}), {\it{i.e.}} the light neutrino mass matrix, via 
\begin{align}
{\mathcal U}_\nu  = 
\begin{pmatrix} 
U_\nu & 0    & 0 \\[0.05cm]
0 & \mathbb{1}  & 0  \\[0.05cm]
0  & 0  & \mathbb{1} 
\end{pmatrix}. 
\end{align} 

\begin{figure}
  \centering
  \begin{tabular}{ccc}
    \includegraphics[width=0.33\textwidth]{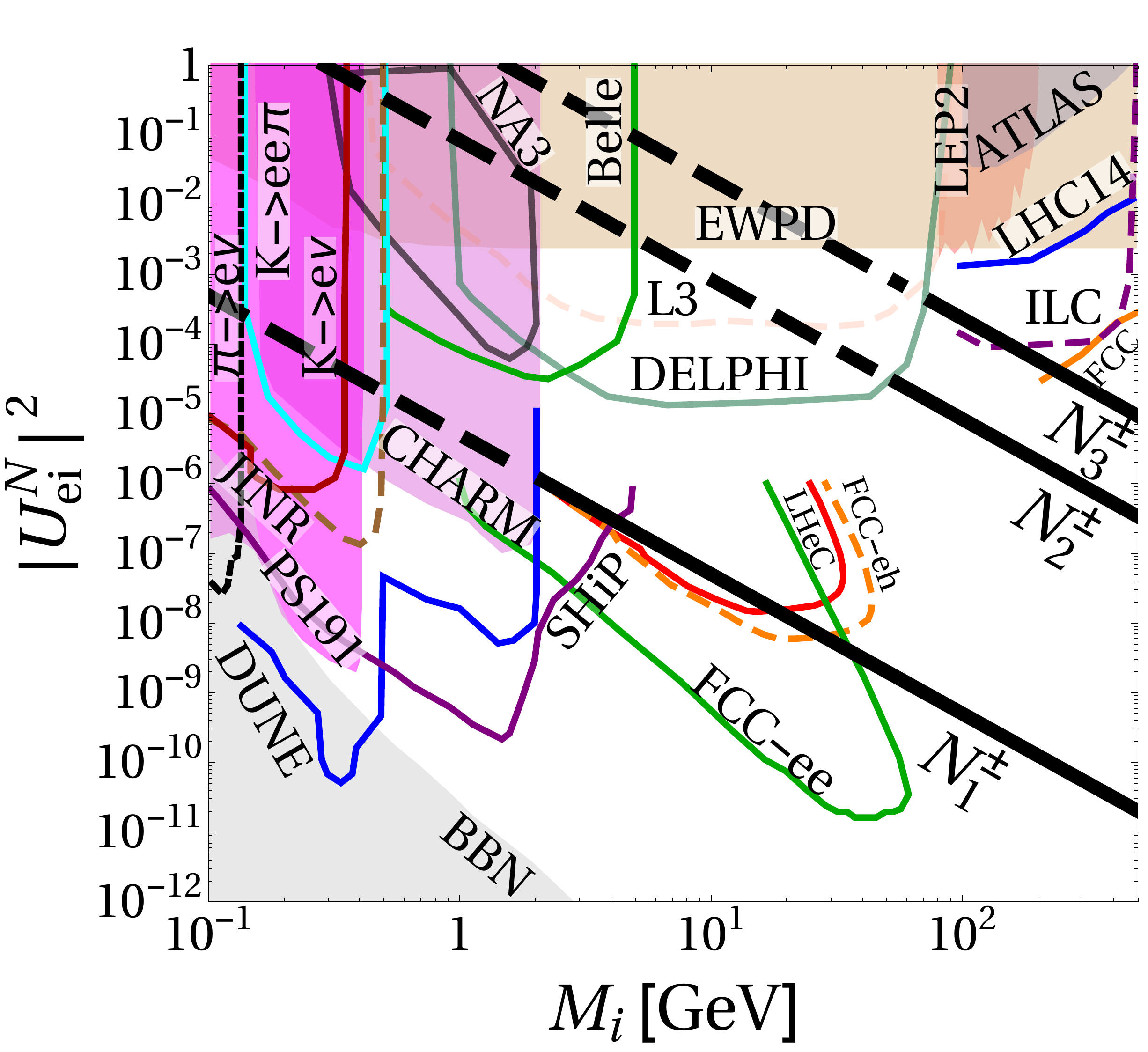} &
    \includegraphics[width=0.33\textwidth]{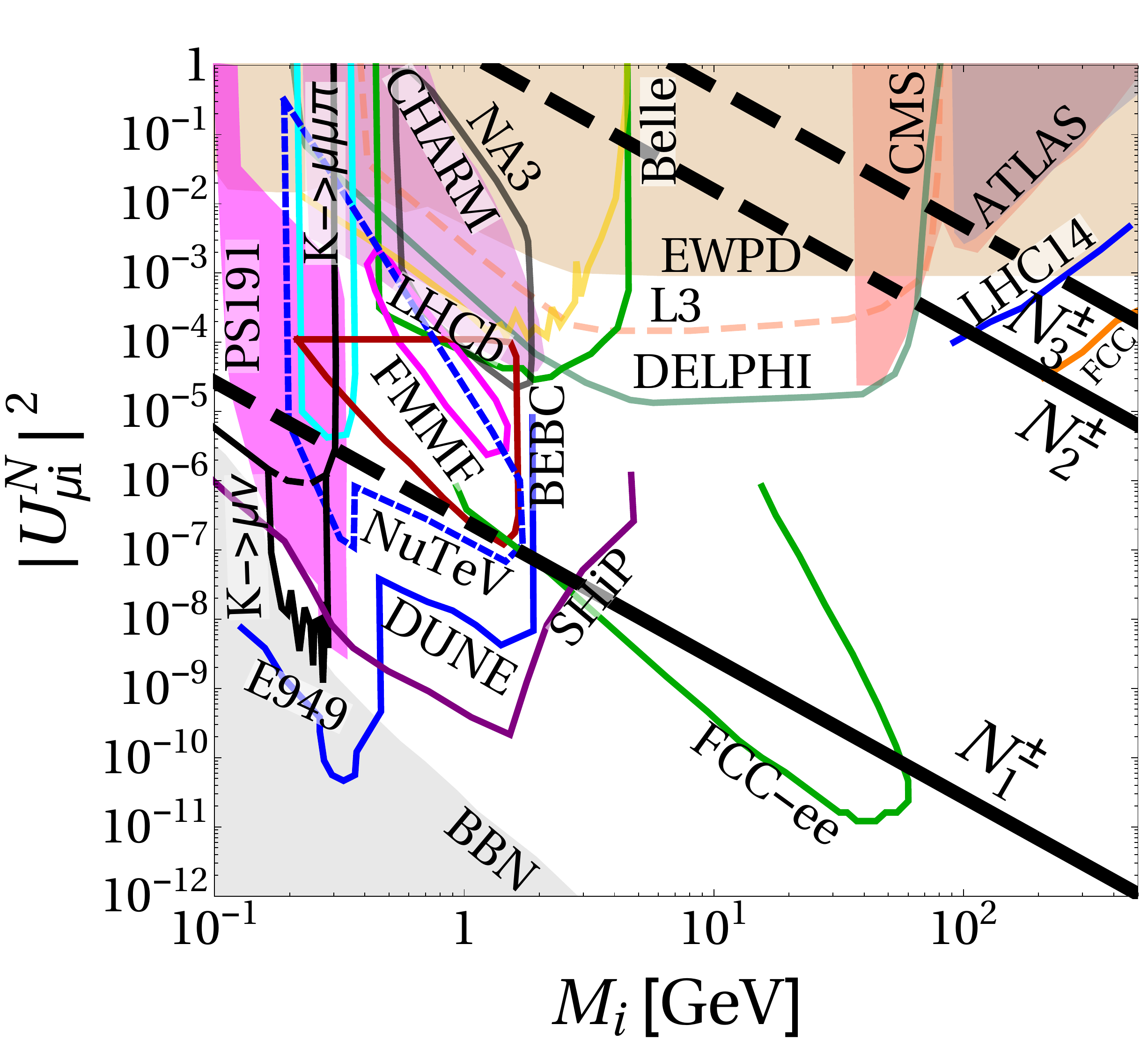} &
    \includegraphics[width=0.33\textwidth]{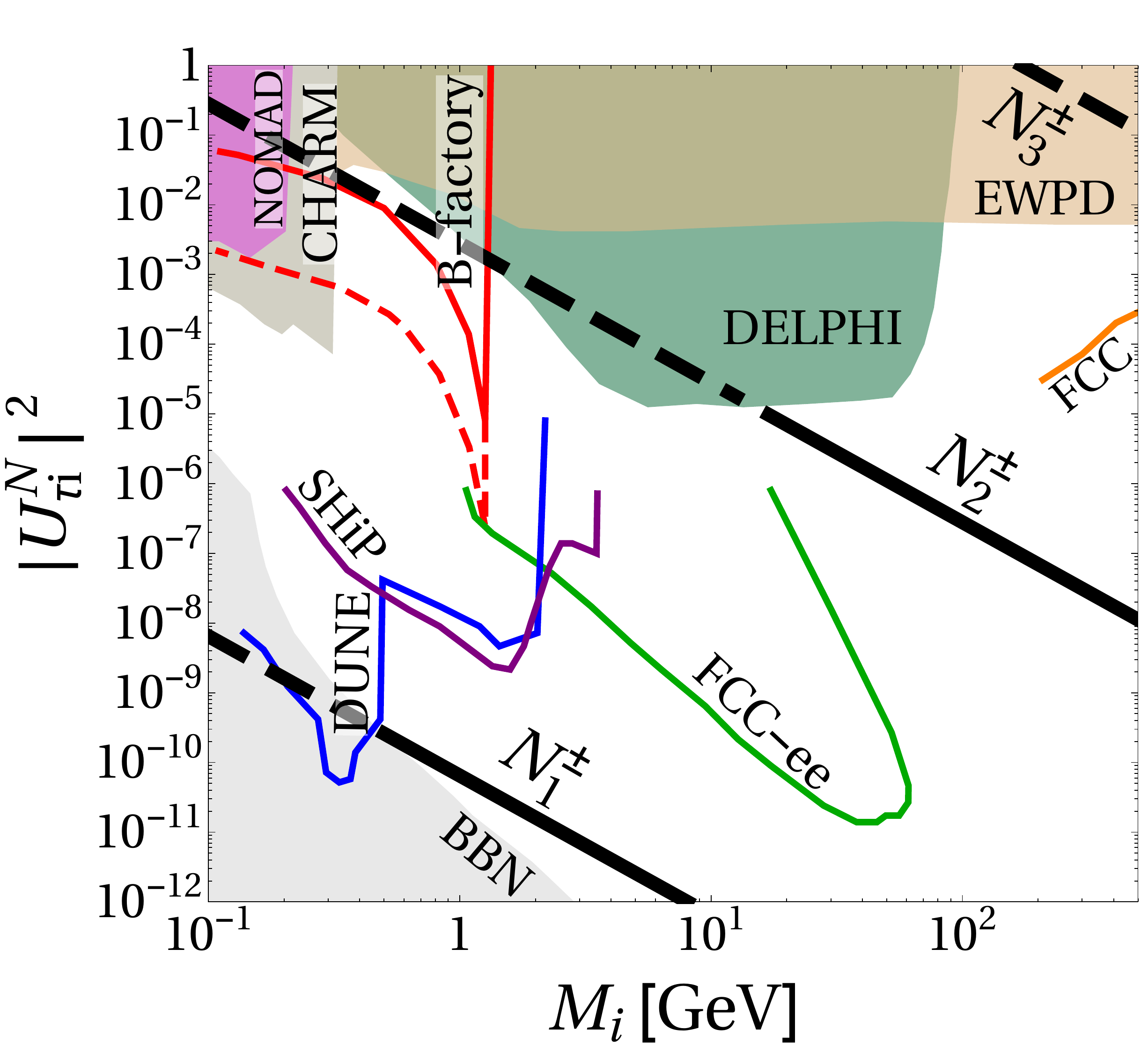} \\
    (a) & (b) & (c)
  \end{tabular}
  \caption{Experimental bounds on and future sensitivities to  the heavy 
lepton mixing in $\nu_e$ (panel (a)), $\nu_\mu$ (panel (b)) and $\nu_\tau$ (panel (c)). 
The black lines show  the predicted mixing of heavy leptons $N_i^\pm \,(i=1,2,3)$ in  a given light neutrino 
flavor state as a function of $M_i$. We take  $(Y_1,\,Y_2,\,Y_3) \simeq \left(2\times 10^{-5},\, 0.007,\,1\right)$. 
Solid (dashed)  parts of the black lines are expectations in the allowed (excluded) region.
}
\label{fig:mixing}
\end{figure}

The transition to the flavor basis requires an additional rotation which diagonalizes the 
mass matrix of the charged leptons 
\begin{align}
{\mathcal U}_l  = 
\begin{pmatrix}
U_l & 0    &     0 \\[0.05cm]
0     & \mathbb{1}  & 0  \\[0.05cm]
0     & 0           & \mathbb{1} 
\end{pmatrix}. 
\end{align}
Then, the total mixing matrix in the flavor basis is given by the product of rotations  
\begin{align}
{\mathcal U}_f  = {\mathcal U}_l^{\dagger}~{\mathcal U}_{S}~{\mathcal U}_{N}~{\mathcal U}_\nu = 
\begin{pmatrix} 
 U_\text{PMNS} & - \vec{s}_N s_\xi\, U_l^\dagger   
& \vec{c}_N s_\xi\, U_l^\dagger \\[0.05cm]
0  & \frac{1}{\sqrt{2}}\mathbb{1}   & \frac{1}{\sqrt{2}}\mathbb{1}  \\[0.05cm]
- s_\xi\, U_\nu   & - \vec{s}_N  \, \mathbb{1} & 
\vec{c}_N  \mathbb{1} \\[0.1cm]
\end{pmatrix}. 
\label{eq:connecting_states_CKM}
\end{align} 
Neglecting $\mu$ contribution in \cref{eq:n_rotationang} and taking $c_N = s_N = 1/\sqrt{2}$ yields
\begin{align}
\nu_\alpha =  U_\text{PMNS}~\nu -\frac{1}{\sqrt{2}} s_\xi\, U_l^\dagger (N^- - N^+)\,,  
\label{eq:activetotal}
\end{align}
where $\nu_\alpha$ and $\nu$ are the flavor and light mass eigenstates, respectively.
Since $U_l$ is  non-diagonal, each active neutrino state 
has admixtures of all pairs of heavy leptons. These admixtures can be constrained by 
various terrestrial experiments as well as by cosmology \cite{Vincent:2014rja}.
In \cref{fig:mixing} we show the bounds (adopted from Ref. \cite{Deppisch:2015qwa}) on the admixtures of $N_i^\pm\,({i=1,2,3})$ in 
$(\nu_e,\,\nu_\mu,\,\nu_\tau)$. Although these bounds have been derived for the mixing 
of a single heavy lepton, they are also applicable 
in our scenario where several heavy states are simultaneously present in the model.

According to \cref{eq:activetotal}, the admixture of $N_i^-$ and $N_i^+$  in $\nu_\alpha$ equals   
\begin{align}
\big|U_{\alpha i}^{N} \big|^2 \equiv \big|U_{\alpha i}^{N^-} \big|^2 =  \big|U_{\alpha i}^{N^+} \big|^2 = 
\frac{1}{2} s_\xi^2\, \big|{U_{l}}_{\,\alpha i}\big|^2 = \frac{1}{2} \left(\frac{{m_{D}}_i}{ M_i}\right)^2 \big|{U_{l}}_{\,\alpha i}\big|^2,   
\label{eq:mixn-alpha}
\end{align}
where in the last equality we expressed $s_\xi$ in terms of $N^\pm$ mass, and we define  $M_i = (M_i^+ - M_i^-)/2$. For relevant cases of our model, the production coherence
of the mass eigenstates  $N_i^+$ and $N_i^-$ is strongly broken,
especially for the lightest leptons ($i = 1$) that appear in low energy
processes. That is, $N_i^+$ and $N_i^-$ are produced (as components of N)
and then decay incoherently without interference effects. Consequently,
equal number of   $l^+$ and $l^-$  leptons will appear in the decays.
For heavier leptons produced in very high energy processes (e.g. decays
of $W_R$)
the coherence can be maintained (see \cite{Das:2017hmg} and
references therein).
The experimental bounds, given in \cref{fig:mixing}, are obtained for a production of a single heavy lepton. 
Since we deal here with two nearly degenerate states that are indistinguishable in experiments, the corresponding bounds on the individual mixing are two times stronger.  
In other words, we can treat the pair as a single particle and multiply the mixing by a factor of two so that the black lines in \cref{fig:mixing} correspond to
\begin{align}
2 |U_{\alpha i}^N|^2 = \frac{m_{Di}^2}{M_i^2} \, |{U_{l}}_{\,\alpha i}|^2\,. 
\end{align}
We use $m_{Di}^2 = ( m_u^2, ~ m_c^2, ~  m_t^2)$, ${U_{l}}_{\,\alpha i} = (V_\text{CKM})_{\alpha i}$ and  do not impose here any relations between the heavy lepton masses $M_i$, in other words we are treating them as independent (the relaxation of the screening condition is discussed in \cref{sec:beyond}).

Notice that since $U_l \sim V_{\text{CKM}} \sim \mathbb{1}$,  the strongest bounds appear in the cases 
when the diagonal elements of $U_l$ are involved. 
The most stringent bound on the mass of $N_1^\pm$ comes from its admixture in $\nu_e$
\begin{align}
\big|U_{e 1}^N\big|^2 \approx \frac{1}{2} \left(\frac{{m_{D}}_1}{M_1}\right)^2 \approx \frac{m_{u}^2}{2 M_1^2}\,, 
\label{eq:1ne-mix}
\end{align}
where $m_{u}\approx(1 - 2)$ MeV is the mass of the up quark at the TeV scale. 
In the left panel of \cref{fig:mixing} we show with a black line (dashed in the excluded parameter space, solid elsewhere) the dependence of $2 \left|U_{e 1}^N\right|^2$  on $M_1$ for $m_{u} \simeq 2$ MeV.   
From this figure we infer 
\begin{align}
M_1 \geq 2 \,\text{GeV}\,,
\label{eq:mass_limit}
\end{align}
which is set by the CHARM \cite{Bergsma:1985is,Orloff:2002de} exclusion region. 
 Varying $m_{u}$ by a factor of $3$ does not change the limit in \cref{eq:mass_limit}. 
However, for $m_u < 0.5$ MeV  the limit becomes weaker: $M_1 \geq 0.4 \,\text{GeV}$.   

Despite the involved CKM suppression, the admixture of $N_1^\pm$ in $\nu_\mu$ yields practically identical (\ref{eq:mass_limit})
bound on $M_1$, which mainly stems from NuTeV \cite{Vaitaitis:1999wq}. From \cref{eq:mixn-alpha} we have 
\begin{align}
\big|U_{\mu 1}^N\big|^2  \approx  \frac{1}{2} \left(\frac{{m_{D}}_1}{M_1^2}\right)^2 \big|{U_{l}}_{\,\mu 1}\big|^2
\approx \frac{m_{u}^2}{2 M_1^2} \sin^2\theta_\text{c}\,, 
\label{eq:1nmu-mix}
\end{align}
where $\theta_\text{c}$ is the Cabibbo angle. 
In turn, this gives the bound 
\begin{align}
\xi \leq 10^{-3}\,.
\label{eq:lim-xi}
\end{align}

In our framework, the neutral lepton masses are related by screening and the $q$-$l$ similarity.
Employing the limit (\ref{eq:mass_limit}) we obtain $M_2 = M_1 m_c/m_u \geq  600$ GeV, where $m_{c} \approx 0.5$ GeV 
is the mass of the charm quark at the TeV scale \cite{Xing:2007fb}. From the screening relation for the mass of the third generation  of heavy leptons, 
$M_3 = M_1 m_t/m_u$, we find the limit on the L-R symmetry breaking scale  
\begin{align}
\langle \chi_R^0 \rangle \approx \sqrt{2}\, M_3  \geq  2 \times 10^5\,\text{GeV}.
\label{eq:lim}
\end{align}

The  condition in \cref{eq:lim} makes discovery of the RH gauge  
bosons and scalars at present colliders unfeasible. 
The leptons  
$N_2^\pm$ and $N_3^\pm$ are beyond the reach of current and future experiments as well 
(see the lines in all panels of \cref{fig:mixing}). 
However, as can be seen from the left and middle panel of \cref{fig:mixing}, 
the lightest pair of neutral leptons is accessible to future colliders \cite{Antusch:2016vyf,dEnterria:2016sca,Blondel:2014bra}, 
beam-dump experiments \cite{Lantwin:2017xtc} and neutrino 
oscillation facilities \cite{Acciarri:2016crz}. In particular, SHiP  
may improve the lower bound (\ref{eq:mass_limit}) on $M_1$ to approximately $5$ GeV, whereas FCC-ee  
will be able to probe even larger masses  (up to $60$ GeV) and very tiny mixing angles. 
Hadron collider FCC-hh with  the total center of mass energy around 100 TeV
will be able to search for $N_2^{\pm}$ (see line denoted {\emph{FCC}} in all panels of \cref{fig:mixing}) and also $N_3^{\pm}$ in the absence of screening \cite{Antusch:2016ejd},
as well as the RH gauge bosons and new scalar bosons from
the Higgs doublets \cite{Golling:2016gvc}.

Without imposing the screening condition, the strongest \emph{direct} 
bound on the mass of $N_2^\pm$ follows from its admixture in $\nu_\mu$ 
\begin{align}
\big|U_{\mu 2}^N\big|^2 \approx \frac{1}{2} \left(\frac{{m_{D}}_2}{M_2}\right)^2 \approx \frac{m_{c}^2}{2 M_2^2}\,. 
\label{eq:2nm-mix}
\end{align}
The regions excluded by DELPHI \cite{Abreu:1996pa} and CMS \cite{Khachatryan:2014dka} experiments give  
$M_2 > 70$ GeV (see $N_2^\pm$ line in the middle panel of \cref{fig:mixing}). 
Decreasing ${m_{D}}_2$ by a factor of $3$ relaxes the bound down to $M_2 > 40$ GeV. 

The admixture of $N_2^\pm$ in $\nu_e$ is characterized by 
\begin{align}
\big|U_{e 2}^N\big|^2 \approx  \frac{1}{2} \left(\frac{{m_{D}}_2}{M_2}\right)^2 \sin^2 \theta_c 
\approx \frac{m_{c}^2}{2 M_2^2} \sin^2 \theta_c\,, 
\end{align}
and according to \cref{fig:mixing} (left panel) this leads to $M_2 > 50$ GeV.  

The strongest \emph{direct} bound on the mass of $N_3^\pm$, $M_3 > 100$ GeV,  
is established by ATLAS. Note that the \emph{direct} bounds on $N_2^\pm$ 
and $N_3^\pm$ are significantly weaker than those obtained 
from the bound on $N_1^\pm$ and the screening relation 
(see \cref{eq:mass_limit,eq:lim}).
Thus, in the absence of screening, the hierarchy of the heavy leptons can be much weaker.   
We will elaborate on such a scenario in \cref{sec:beyond}. 

Let us note that the neutral leptons, $(N^+ + N^-)/\sqrt{2}$, are produced dominantly 
via the mixing with active neutrinos in the processes involving the left-handed
gauge bosons \cite{Banerjee:2015gca}. The production of $N^-$ and $N^+$ via the exchange of the off-shell right-handed bosons in $e^+ e^-$, $p\bar{p}$ and $pp$ collisions is subdominant 
as the corresponding cross-sections are suppressed by a factor 
$\xi^4=\left(\langle \phi_1^0 \rangle / \langle \chi_R^0 \rangle\right)^4$.

\subsection{$0\nu2\beta$ Decay}
\label{subsec:0nu2b}
\noindent
The dominant contribution to the neutrinoless double beta ($0\nu 2\beta$) decay arises 
from the left-handed current since the right-handed current contribution 
scales as $\xi^4 \lesssim 10^{-12}$ \cite{Dev:2013vxa,Pas:2015eia,Deppisch:2017vne} 
due to the bound (\ref{eq:lim-xi}). 
Then, the effective Majorana mass of the electron neutrino can be presented as
\begin{align}
m_{ee} = m_{ee}^l + m_{ee}^h\,, 
\end{align}
where the contributions from the light  and heavy mass eigenstates read 
\begin{align}
 m_{ee}^l   &=  \,\sum_{i = 1,2,3} (U_{\text{PMNS}\,ei})^2 \,m_i  \approx s_\xi^2\, \mu_{ee}\,,\nonumber\\ 
m_{ee}^h &= r\, p^2 \sum_{i = 1,2,3} \left( (U_{ei}^{N^-})^2 \frac{M_i}{p^2 - M_i^2} + 
(U_{ei}^{N^+})^2 \frac{M_i'}{p^2 - {M_i'}^2} \right) \equiv \sum_{i = 1,2,3} m_{i}^h\,.  
\label{eq:eelight+heavy}
\end{align}
Here, $r \equiv M^l_{\beta\beta 0\nu}/M^h_{\beta\beta 0\nu}$ is the ratio of nuclear matrix elements 
for the exchange of heavy and light neutrinos, $- p^2 \sim (125~{\rm MeV})^2$ is the neutrino momentum squared and $\mu_{ee}$ denotes the $(1,1)$ element of the $\mu$ matrix. 

The contribution from the $i$-th pair of pseudo-Dirac
neutrinos ($N_i^-$ and $N_i^+$) can be estimated as
\begin{equation}
m_i^h = \xi^2 \mu_{ee} \frac{p^2}{M_i^2}
\sim m_{ee}^{l} \frac{p^2}{M_i^2}\,,
\label{eq:eeheavyes}
\end{equation}
where  $\xi^2$ arises from  the admixture of $N_i^\pm$ in $\nu_e$
and $\mu_{ee}$ stems from the sum $M_i^+ + M_i^-$ (see \cref{eq:massMM}).  
Due to strong mass hierarchy, the contributions from the heavier pairs 
are negligible and only the lightest one ($N_1^-$ and $N_1^+$) should be considered. Hence, $m_{ee}^h \approx m_1^h$.

To compute $m_1^h$ one should retain the $1$-$2$ and $1$-$3$
blocks in the matrix (\ref{eq:mass}) and  perform further rotations to diagonalize it up to $\mathcal{O}(\mu^2)$. 
The terms proportional to the elements of $\mu$ matrix can not be neglected and, in particular, deviations of the $N_L-S$ mixing from $45^\circ$ 
should be taken into account in  \cref{eq:connecting_states_CKM}. 
To diagonalize the matrix in \cref{eq:mass} we perform two additional
rotations:
\begin{align}
{\mathcal U}_{14}  = 
\begin{pmatrix} 
\mathbb{1}   & U_{14}  & 0  \\[0.05cm]
    -U_{14} &  \mathbb{1} & 0 \\[0.05cm]
   0  &   0   & \mathbb{1} 
\end{pmatrix}, ~~~~~
{\mathcal U}_{17} = 
\begin{pmatrix} 
  \mathbb{1}   & 0  & U_{17} \\[0.05cm]
    0  &  \mathbb{1}  & 0  \\[0.05cm]
   - U_{17}  &   0  & \mathbb{1}  
\end{pmatrix},
\label{eq:matrices_0nu2beta}
\end{align} 
with 
\begin{align}
U_{14} \approx U_{17} =\begin{pmatrix} 
  -\frac{s_\xi c_\xi}{\sqrt{2}} \frac{\mu_{ee}}{M_1}   & 0   & 0 \\[0.05cm]
  0 &   0   &    0  \\[0.05cm]
 0 &  0   &  0
\end{pmatrix}. 
\end{align}
\noindent
Now the total mixing matrix equals
\begin{align}
{\mathcal U}_f  = {\mathcal U}_l^{\dagger}~{\mathcal U}_{S}~{\mathcal U}_{N}~{\mathcal U}_\nu~ {\mathcal U}_{14}~
{\mathcal U}_{17}\,.
\label{eq:totalmix}
\end{align}
According to  \cref{eq:totalmix}, the flavor neutrino states can be expressed in terms of mass eigenstates 
\begin{equation}
\nu_f = c_\xi\, U_{\text{PMNS}}\, \nu +
\left(c_\xi\, U_{\text{PMNS}}\, U_{14} - s_\xi\, s_N\, U_{l}^\dagger
\right)\,N_1^- +
\left(c_\xi\, U_{\text{PMNS}}\, U_{17} + s_\xi\, c_N\, U_{l}^\dagger
\right)\,N_1^+\,,
\label{eq:flavn1}
\end{equation}
where we neglected the small correction to the first term.
Then, the admixtures of $N_1^-$ and $N_1^+$ in $\nu_e$ are given explicitly  by
\begin{align}
\mp \frac{s_\xi}{\sqrt{2}}
\left[1  \pm  \frac{c_\xi^2 \mu_{ee}}{M_1} 
\left( (U_{\text{PMNS}})_{e1}  - \frac{1}{4} (U_l^\dagger)_{e1} \right) 
\right]\,,
\label{eq:eN1mix}
\end{align}
where the upper (lower) sign corresponds to $U_{e1}^{N^-}$ $(U_{e1}^{N^+})$.
After inserting these expressions into \cref{eq:eelight+heavy}, we find
\begin{equation}
m_{1}^{h} = - 2 \,(U_{\text{PMNS}})_{e1}\,(U_l^\dagger)_{e1}\,
\frac{|p^2|}{M_1^2}\, s_\xi^2\, c_\xi^2\, \mu_{ee}
=  - 2 (U_{\text{PMNS}})_{e1}\,(U_l^\dagger)_{e1}\,
\frac{|p^2|}{M_1^2}\, m_{ee}^l\,. 
\label{eq:meeheavy1}
\end{equation}
Thus, the mass $m_{1}^{h}$  is proportional to the contribution from light neutrinos, $m_{ee}^l$,
with the additional suppression factor $|p^2|/M_1^2 $, in agreement with the estimate in  \cref{eq:eeheavyes}.
The ratio of the two contributions equals 
\begin{equation}
\frac{m_{1}^{h}}{m_{ee}^l} = 1.67 \,\frac{|p^2|}{M_1^2}\,.
\label{eq:meeheavy2}
\end{equation}
For $M_1 = 0.4$ GeV and $M_1=2$ GeV this suppression is $0.1$ and $0.004$, respectively. 
Hence, the heavy lepton contribution is practically not observable in the $0\nu2\beta$ decay experiments. 

\subsection{Leptogenesis}
\label{subsec:lepto}
\noindent
The inverse seesaw mechanism  with its pairs of quasi-degenerate
heavy leptons appears, at a first sight, very suitable for realization of
low scale leptogenesis mechanisms. Those include the  resonant leptogenesis
 \cite{Pilaftsis:2003gt,DeSimone:2007edo}, where the lepton
asymmetry is produced from the decays of quasi-degenerate heavy leptons, and
the ARS mechanism based on  oscillations between
heavy neutral leptons \cite{Akhmedov:1998qx}.
A vast majority of the previous studies on leptogenesis in the inverse seesaw framework have been performed in the SM gauge structure. In the  case of  L-R model with the $q$-$l$ similarity
we have large Yukawa couplings and additional processes which enhance the washout effect.

In its minimal setup, the inverse seesaw model is not compatible
with successful thermal leptogenesis due to strong washout \cite{Dev:2014laa,Dev:2017wwc}.
The maximal achievable value of the baryon asymmetry is $8$ orders of magnitude smaller than the observed one \cite{Agashe:2018oyk}.

In our L-R symmetric setup, similar conclusions  apply.
Furthermore,  what already prevents any possibility for viable leptogenesis
in our model is the size of the Yukawa couplings chosen according to the $q$-$l$ similarity.
As a result, the out-of-equilibrium condition is not satisfied
for all pairs of heavy leptons.  In particular, for the second pair with $M_2\sim$ TeV
there is no deviation from thermal equilibrium for couplings $Y_{2} \gtrsim \mathcal{O}(10^{-7})$.  For $Y_{2}\lesssim \mathcal{O}(10^{-7})$
we obtain $\left(N^\pm-N^{eq}\right)/N_{eq} \gtrsim 10^{-1}$  at $M/T \gtrsim \mathcal{O}(1)$. This should be compared with  $Y_2 \sim 5\cdot 10^{-3}$ in our model.

The observed baryon asymmetry can be produced in the non-minimal realization of the inverse seesaw \cite{Agashe:2018oyk,Aoki:2015owa} which is  motivated
by explanation of the scale of Majorana masses $\mu$. The extension includes  an extra fermion singlet $X$ with large  Majorana mass $M_X \gg M_i$ and the Higgs singlet $\sigma$ which develops a VEV. The authors \cite{Agashe:2018oyk} employ the additional global symmetry under which $S$ is charged. This is, however, not possible in our scenario since $S$, $\sigma$ and $X$ are not charged under any global symmetry.



The leptogenesis mechanisms discussed so far either fail in producing the required amount 
of asymmetry or are not compatible with our model setup.  
For alternative and potentially successful low-scale leptogenesis scenarios (see below) it is necessary to avoid strong washout.  
At $T\sim M_N$, the lepton number violating process $ L_L \Phi \to L_L^c \Phi^\dag$ mediated by the Majorana fermions $N^-$ and $N^+$ becomes operative. The strength of the washout is quantified by the parameter $K_i$ which, for a given fermion generation $i$, reads

\begin{subequations}
  \label{eq:all}
    \begin{empheq}[left={K_i=\empheqlbrace\,}]{align}
      & \frac{\Gamma_i}{H(T=M_i)}  \left(\frac{\mu_i}{\Gamma_i} \right)^2 \quad,\,\,\text{for}\,\,
   \mu\ll \Gamma_i\,,
        \label{eq:washout_1} \\
      & \frac{\Gamma_i}{H(T=M_i)} \hspace{1.7cm},\,\,\text{for}\,\, \mu \geq \Gamma_i\,\,.
       \label{eq:washout_2}
    \end{empheq}
\end{subequations}
Here, $\Gamma_i= Y_i^2 M_i /(8\pi)$
is the decay rate of $N_i^\pm$.
The additional factor  $(\mu_i / \Gamma_i)^2$ in \cref{eq:washout_1} originates from the interference
between diagrams containing quasi-degenerate particles ($N^-$ and $N^+$) \cite{Blanchet:2009kk,Blanchet:2010kw}.

Let us make general estimations by dropping the assumptions of the $q$-$l$ similarity and screening.
From the inverse seesaw formula we have  $\mu = 2\,m_{\nu}\, M^2/(Y^2 \langle  \phi_1^0\rangle^2)$,  
and  consequently
\begin{align}
\frac{\mu}{\Gamma}  =  \frac{16\pi \, m_\nu}{\langle  \phi_1^0\rangle^2} \frac{M}{Y^4} =
8.3 \cdot 10^{-9}\,\frac{1}{Y^4} \left(\frac{m_\nu}{ 0.1 {\rm~ eV}}\right) \left(\frac{M}{10^5\,\text{GeV}}\right),
\label{eq:mugamma}
\end{align}
where for brevity we omitted flavor generation indices.
For $m_\nu = 0.01$ eV, $M = 200$ GeV and $Y = 1.1 \cdot 10^{-3}$, we obtain $\mu/\Gamma = 1$. For smaller $Y$, we should use expression (\ref{eq:washout_2})  which gives $K \gg 1$, implying a very strong washout unless $Y\lesssim\mathcal{O}(10^{-7})$.
For $Y > 1.1 \cdot 10^{-3}$, the expression (\ref{eq:washout_1}) should be used. It can be rewritten as
\begin{align}
K =
{32\pi} \frac{m_\nu^2 M}{Y^6 a \langle  \phi_1^0\rangle^4} \simeq
 1.95\cdot 10^{-5}  \frac{1}{Y^6}  \left(\frac{M}{10^5 ~{\rm GeV}}\right)
\left(\frac{m_\nu}{ 0.1 ~{\rm eV}}\right)^2,
\label{eq:washout_1a}
\end{align}
where $a = 1.66 \sqrt{g_*} /M_{\text{Pl}}$. The condition $K < 1$ gives even stronger bound on the coupling, namely
$Y > 0.06$ for $M > 200$ GeV. Together with the bound on the admixture of the heavy leptons in the flavor states
$\xi = Y v_L /M < 10^{-2}$ (which is not applicable above TeV scale) this leads to a narrow range of the allowed parameters:
$M = (10^3 - 10^5)$ GeV and $Y > 0.1$, which is satisfied only for the third generation.

For the inverse seesaw, the ARS mechanism leads to successful leptogenesis
if $Y \sim 10^{-7}-10^{-6}$ \cite{Abada:2017ieq}. 
For masses of $N^-$ and $N^+$ at $\mathcal{O}(\text{TeV})$ the washout (\ref{eq:washout_2}) is suppressed. Note, however, that such parameter space is not in accord with $q-l$ similarity. 

One can rely on the electroweak baryogenesis \cite{Morrissey:2012db} which requires
a strong first order electroweak phase transition. The rich scalar sector
in our  model may help  to realize such a scenario. Still, the lepton number washout should be suppressed, since sphalerons partially carry the baryon asymmetry to the lepton section.

\subsection{Corrections to the Higgs mass. Naturalness}
\label{subsec:higgs}
\noindent

In this section we will address a specific
problem related to existence of heavy RH neutrinos (recall that for the
third neutrino $M_3 \gg v_{EW}$)
and their radiative corrections to the Higgs mass.
It was noticed long time ago that the contribution to the Higgs mass from the loops 
formed by active neutrinos and heavy RH neutrinos in the type I seesaw mechanism equals \cite{Vissani:1997ys} (see \cite{Bazzocchi:2012de,Clarke:2015gwa,Bambhaniya:2016rbb} for recent studies)
\begin{align}
\delta m_H^2 \approx  \frac{Y_i^2 M_i^2}{4\pi^2} = 
\frac{m_{\nu 3} M_3^3}{2\pi^2\,\langle \phi_1^0\rangle^2}\,.    
\label{eq:seesaw1}
\end{align}
Then, the  condition that the correction $\delta m_H^2$ (\ref{eq:seesaw1}) does not exceed 
the Higgs mass itself (``naturalness''), $\delta {m_H^2}\lesssim 10^4 \,\text{GeV}^2$, leads to the upper bound 
on the mass  \cite{Vissani:1997ys}  
\begin{align}
M_i \lesssim \,\text{few}\times 10^7\, \text{GeV}\,,
\label{eq:bound_Vissani}
\end{align}
where  $m_{\nu 3}\sim 0.1$ eV  was used. This is smaller than the standard lower
bound from thermal leptogenesis \cite{Davidson:2002qv}  (however, see \cite{Moffat:2018wke}).

With $M_i \leq 10^{5}$ GeV (that we obtained in \cref{subsec:sterile}), the bound (\ref{eq:bound_Vissani}) is satisfied.
However, in the inverse seesaw, the correction to $\delta m_H^2$ becomes larger
than in the type I seesaw model, being enhanced by the factor  $M_i/\mu$  \cite{Haba:2016lxc}
\begin{align}
\delta m_H^2 \sim \frac{m_{\nu 3} \,M_i^4}{2\pi^2\, \mu_{max}\,\langle \phi_1^0\rangle^2}\,.
\label{eq:iss}
\end{align}
Here, $\mu_{max}$ represents the largest entry of the $\mu$ matrix. 
The difference from expression (\ref{eq:seesaw1}) originates from different dependence of the neutrino mass on the mass of 
heavy leptons. \cref{eq:iss} and the inequality  $\delta {m_H^2} < 10^4 \,\text{GeV}^2$
give the bound $M_i < 10^4$ GeV for $\mu_\text{max}\sim\mathcal{O}(100)$ keV. This bound is  about 1 order of magnitude lower than  $M_3$. 
At the same time, in our scenario (in contrast to, e.g., Ref.  \cite{Vissani:1997ys}), there are
other particles, gauge bosons and scalars, at the scale $10^5$ GeV.
Being bosons,
these particles  give corrections to the Higgs mass of the opposite sign
and they can cancel contributions from the RH neutrinos.
Let us consider such a possibility.

For the gauge boson contribution we consider the 1-loop corrections
with heavy right-handed charged ($W_R^\pm$) and neutral ($Z_R$) vector bosons.
To estimate $\delta {m_H^2}$ arising  from the loops of $W_R^\pm$, $Z_R$ and $N_3^\pm$ we use the 1-loop effective potential \cite{CW, Casas:1999cd}.
Expanding the 1-loop potential and identifying the term containing two SM Higgs fields
we obtain the condition required for a  1-loop cancellation between the fermion and gauge boson contributions to $\delta {m_H^2}$
\begin{align}
\frac{Y_3^2}{g^2}\simeq   \left (\frac{11}{8}\right)^{1/2} \approx 1.17\,.
\label{eq:f+g}
\end{align}
In \cref{eq:f+g}, $g$ is the SU(2) gauge coupling.
In our framework, Yukawa couplings are fixed according to the $q$-$l$ similarity 
condition  and the gauge coupling for $SU(2)_R$ equals the $SU(2)_L$ one. The values of $g$ and $Y_3$ at the top quark mass scale are $0.65$ and 
 $0.93$, respectively, so that  ${Y_3^2}/{g^2} = 2.05$ which clearly fails to satisfy \cref{eq:f+g}. 
 Even though the renormalization group effects may help a bit (we find 7\%
decrease of $Y_3/g$ at $\langle \chi_R^0\rangle \sim 10^5$\,GeV) 
the cancellation can not be achieved without additional scalar contribution. Another possibility is that $Y_3$ differs from the top Yukawa coupling.  

In the scalar sector a number of diagrams can contribute to the SM Higgs mass. If only one scalar with mass $m^2\sim (c/2) \,\langle \chi_R^0 \rangle^2$ gives the dominant contribution to $\delta {m_H^2}$, the condition for complete cancellation of bosonic and fermionic 1-loop contributions to the SM Higgs mass reads
\begin{align}
11 \,g^4 - 8 \,Y_3^4 +2 \,c^2 \simeq 0\,. 
\label{eq:f+g+s}
\end{align}
Note that for this cancellation one needs specific values of couplings in the scalar sector. There is no principle which would fix the couplings to such values implying that if the cancellation occurs it is accidental.
If the full 1-loop cancellation is achieved, the dominant Higgs mass correction stems from 2-loop diagrams 
which can be estimated as \cite{Fabbrichesi:2015zna}
\begin{align}
\delta m_H^2 \sim \frac{g^2 Y_3^2}{(16 \pi^2)^2} M_3^2  \approx \frac{g^2}{2(16 \pi^2)^2} \langle \chi_R^0 \rangle^2  = 3.4\cdot 10^5 \,\text{GeV}^2\,.
\label{eq:2loop}
\end{align}
 The value given in \cref{eq:2loop} is only one order of magnitude larger than  
$m_H^2$, and therefore a rather acceptable level of fine-tuning is required. Note again that a small deviation from the  $q$-$l$ similarity can also solve the problem. 

Let us underline that in our scenario new physics scale is just 3 orders
of magnitude larger than the EW scale and hence we deal here with mild hierarchy in the scalar sector.
We do not discuss the origin of this hierarchy, but we find that the
hierarchy is supported by the price of moderate fine tuning. The hierarchy can be further weakened if we depart from the q-l similarity.

Of course, in the complete analysis one should take into account all
corrections to the Higgs mass including corrections from the top quark. Note that in the dimensional regularization the leading
corrections are proportional to the mass of particle propagating in the loop.
Therefore, the correction from top and
other EW scale particles are much smaller that those from heavy right-handed
neutrinos.

\section{Variation on the theme}
\label{sec:beyond}

\subsection{Altering heavy fermion portal couplings}
\label{subsec:altering}
\noindent
In the scenario described in \cref{subsec:sterile}
the bound on the scale of L-R symmetry breaking has been obtained using the following three points: (\emph{i}) the lower bound on the mass of the lightest heavy leptons $N_1^\pm$ (see \cref{eq:mass_limit}), 
(\emph{ii}) the screening (\ref{eq:scren}), 
and (\emph{iii}) the $q$-$l$ similarity, \cref{eq:equal}.
The latter implies the equalities $M_3 = M_1  m_t/m_u$ and  $Y_3 = 1$. Consequently,
\begin{equation}
\langle \chi_R^0 \rangle \simeq \sqrt{2} \, M_3 = \sqrt{2}\, M_1 \,\frac{m_t}{m_u}.
\label{eq:rightscale}
\end{equation}
The scale of $\langle \chi_R^0 \rangle$ can be reduced if the assumptions in  
\cref{eq:scren} and/or \cref{eq:equal} are relaxed.
Let us discuss the two scenarios in which  one of these assumptions is abandoned.\\

\noindent
$(i)$ \emph{Departing from the $q$-$l$ similarity}\\

We keep the screening which can be expressed as 
\begin{equation}
Y_i = R_0 \,Y_{Ri}, 
\label{eq:screengen}
\end{equation}
where in general $R_0 \neq 1$. This implies  
\begin{equation}
\xi_i \equiv \frac{{m_{D}}_{i}}{{M_{D}}_{i}} =     
\frac{\langle \phi_1^0 \rangle}{\langle \chi_R^0 \rangle} 
\frac{Y_{i}}{Y_{Ri}} = 
\frac{\langle \phi_1^0 \rangle}{\langle \chi_R^0 \rangle}R_0  
\equiv \xi, 
\end{equation}
{\it i.e.} all $\xi_i$ are equal and \cref{eq:lightmm} for $m_\nu$ is unchanged. 
Let us recall that $\xi$ determines admixtures of heavy leptons in the flavor states. 
In the absence of the $q$-$l$ similarity, the Dirac masses ${m_{D}}_i$ are free parameters. 

The strongest bound on $\xi$ is obtained from mixing 
of $N_1^\pm$ in $\nu_e$ since $\xi^2 = |U_{e1}^N|^2$. From \cref{fig:mixing} one finds   
$\xi \leq 0.006$ and  $M_1 > 2$ GeV. This corresponds to  
$m_{D1} \leq 12$ MeV. 

As the Dirac masses ${m_{D}}_i$ are not fixed, even the extreme scenario
with all ${m_{D}}_i$ (and therefore $M_i$)  being equal is possible.
For $m_{Di} = 12$ MeV  all the heavy leptons are accessible to  experiments.
From the expression for the light neutrino masses (\ref{eq:lightmm}), 
we find  that for $\xi = 0.006$ and $m_{\nu 3} = 0.05$ eV  
the entries of the Majorana mass $\mu$ should be $\mathcal{O}(1)$ keV. 

The scale of L-R symmetry breaking can be substantially lowered in comparison to the value given in \cref{eq:lim}. 
Namely, the bound can be as low as the bound from the direct RH gauge boson searches, which is roughly $3$ TeV \cite{Khachatryan:2014dka,Patra:2015bga}.\\

\noindent
$(ii)$ \emph{Departing from the exact screening}\\

Now we keep the $q$-$l$ similarity only. 
In this case, $\xi_i \equiv {m_{D}}_{i}/{M_{D}}_{i}$ are different. 
Taking weaker  hierarchy of $M_i$ than that of $m_{Di}$, for instance $M_i = (2, ~200 ,~ 5 \cdot 10^{3})$ GeV, 
we have  $\xi_i = (1.15\cdot 10^{-3},~~ 6.45 \cdot 10^{-3},~~ 3.45 \cdot 10^{-2})$.
The corresponding effective mixing parameters given by $|\xi_i|^2$ yield
\begin{align}
2 |U_{e 1}^N|^2  &\simeq 1.32 \cdot 10^{-6}\,,  & 2 |U_{\mu 2}^N|^2  &\simeq 4.16 \cdot 10^{-5}\,,  &
2 |U_{\tau 3}^N|^2  &\simeq 1.18 \cdot 10^{-3}\,. 
\end{align}
Now,  $\langle \chi_R^0\rangle = \sqrt{2}\,M_3 \simeq 7$ TeV for ${Y_{R}}_3\simeq 1$,
so that both additional neutral leptons and right-handed gauge bosons are accessible to LHC.

Due to departure from screening, the formula for light neutrino masses changes
\begin{align}
m_\nu &= \left(\frac{\langle \phi_1^0 \rangle}{\langle \chi_R^0 \rangle}\right)^2 R \,\mu\, R\,, &
R &\equiv \text{diag}\left(Y_1/{Y_{R}}_1, ~Y_2/{Y_{R}}_2,~ Y_3/{Y_{R}}_3\right),
\label{eq:massnoscr}
\end{align}
where $R$ can be absorbed in redefinition of the matrix $\mu$:
\begin{equation}
\mu_{ij} \rightarrow \mu_{ij}' = \mu_{ij} \,R_i \, R_j\,. 
\label{eq:didj}
\end{equation}
Numerically, for chosen $M_i$
we obtain $R \simeq (0.033, ~ 0.185, ~ 1)$. 

The appearance of $R$ may, however, complicate the explanation
of mixing pattern from symmetry arguments since now both $Y$ and $Y_R$ are non-trivially 
involved in the expression for the light neutrino mass matrix. 
Consequently, certain correlation between $R$ and $\mu$ matrices should exist. 
Also, $R$ would affect phenomenology of the heavy leptons (production, decay,  {\it{etc.}}).

\subsection{Left and Right fermion singlets}
\noindent

So far we considered the scenario with single Majorana fermion $S$ per generation, that is the common 
fermion $S$ for the left and right sectors.   
This is consistent with L-R symmetry. Under $P$ transformations we had $S_L \leftrightarrow (S^c)_R$. 
Minimal and logically straighforward extension of this scenario is a P-symmetric 
model with two independent singlets $S_L$ and $S_R$ 
for the left and the right sectors, respectively. 
This study allows to check wheather the L-R symmetry can be realized in the singlet sector.

Now the Yukawa interactions and mass terms read
\begin{align}
{\mathcal L} \supset
- &\bar{L}_R \,Y\, \Phi^\dagger L_L - 
\bar{L}_R \,\tilde{Y}\, \tilde\Phi^\dagger L_L
- \bar{S}_L^c \,Y_L\, \tilde{\chi}_L^\dagger L_L
- \bar{S}_R^c\, Y_R\, \tilde{\chi}_R^\dagger\, L_R \,-
 \nonumber \\& \frac{1}{2} \bigg[\bar{S}_L^c \,\mu_{LL}\, S_L
+ \bar{S}_R^c \,\mu_{RR}\, S_R
+ \bar{S}_L \,\mu_{LR}\, S_R\bigg] + \text{h.c.}\,.
\label{eq:addlagr}
\end{align}
Due to P-invariance, the relations (\ref{eq:parity_rel}) involving $Y$ and $Y_{L(R)}$ still hold as before. 
Now, the Yukawa interactions in (\ref{eq:addlagr}) are invariant with respect to  global
$U(1)$ symmetry  of the lepton number with charge prescription
$L_g(L_L) = L_g(L_R) = 1$, $L_g(S_L) = L_g(S_R) = - 1$, and zero charges for scalar fields. 
This symmetry forbids the Yukawa interactions  $\bar{S}_{R(L)} \ \tilde{\chi}_{L(R)}^\dagger L_{L(R)}$. 
The symmetry is broken by the Majorana mass terms
(second line in \cref{eq:addlagr}). The smallness of masses  $\mu$ can be related to this breaking.

Invariance with respect to $P$ transformation $S_L \leftrightarrow  S_R$ would imply the following equalities:
$\mu_{LL}=\mu_{RR}, \, \mu_{LR}=\mu_{LR}^\dagger\,$. 
However,  as in the visible sector, masses can break parity,  so that
in general $\mu_{LL} \neq \mu_{RR}$.
The breaking can be spontaneous if $\mu$ terms are generated
by couplings of $S$ with singlet scalars $\sigma_L$, $\sigma_R$
$\sigma_{LR}$: $y_L \bar{S}_L^c S_L \sigma_L$,
$y_R \bar{S}_R^c S_R \sigma_R$. Then, even if we impose
P-symmetry which gives $y_R = y_L$, the scales of
$\mu_{LL}$ and $\mu_{RR}$ can be different due to
$\langle \sigma_L \rangle \neq \langle \sigma_R \rangle$,
i.e.  spontaneous violation of parity in the singlet sector.
In this case $\mu_{LL} \propto \mu_{RR}$. 
With complicated singlet sector one can obtain also different
structures of matrices $\mu_{LL}$ and $\mu_{RR}$.
Finally, the L-R symmetry may be explicitly broken in
the S-sector. In what follows we will not specify origins of $\mu$ matrices,
but consider {\it a priori} general structure of $\mu_{LL}$ and $\mu_{LR}$
assuming only that $\mu \ll v_{EW}$. As before,  $\mu_{RR}$ is fixed
via inverse seesaw by masses and mixing of light active neutrinos.
Notice that in this extension the  contribution to the $\chi_L^{\dagger} \tilde{\Phi} \chi_R$ coupling 
is generated by the loop diagram of \cref{fig:vev} with $\mu$ substituted by $\mu_{LR}$.

After the Higgs fields acquire VEVs, the mass matrix in the $(\nu_L, ~ N_L, ~ S_L, ~ S_R^c)$ basis reads
\begin{align}
{\mathcal M} = 
\begin{pmatrix} 
  0 & m_D & m_D' & 0  \\[0.1cm]
  m_D & 0 & 0  &  M_D     \\[0.1cm]
  m_D' & 0 & \mu_{LL}  &  \mu_{LR}  \\[0.1cm]
   0 & M_D & \mu_{LR}^T  &  \mu_{RR} 
\end{pmatrix},
\label{eq:connecting_states}
\end{align} 
where all entries are $3\times 3$ matrices and the expressions for $m_D$, $m_D'$ and 
$M_D$ are given in \cref{eq:r}.
The procedure of diagonalization of this matrix is similar to the one outlined in \cref{subsec:sterile}.  
First, we make a rotation in the $\nu_L - S_R^c$ plane by $s_\xi$.
In the new basis $(\nu', ~ N_L, ~ S_L, ~ S')$ we perform nearly maximal rotation in the
$(N_L - S')$ plane. Then, in the rotated basis $(\nu', ~ N^-, ~ S_L, ~ N^+)$, 
$N^-$ and $S_L$ are permuted and the pseudo-Dirac states ($N^-$ and $N^+$) decouple.  
In the basis of light states $(\nu', S_L)$ the mass matrix reads
\begin{align}
\begin{pmatrix} 
  \mu_{RR} \, s^2_\xi & c_\xi\, m_D' - s_\xi \,\mu_{LR} \\[0.1cm]
  c_\xi \, m_D' -  s_\xi\,\mu_{LR}  & \mu_{LL}
\end{pmatrix}\,.
\label{eq:2states}
\end{align}
Interestingly, the decoupling of $N^-$ and $N^+$ does not produce
 $\mathcal{O}(\mu_{ij}/{M_D})$ corrections to this matrix.

Phenomenology of this extended version is largely identical to the one  of
the main scenario. In particular, properties of heavy pseudo-Dirac states
formed now by $\nu_R$ and $S_R$ are similar to ones presented in \cref{subsec:sterile}. 
The light neutrino mass matrix is given by the $(1,1)$ element of (\ref{eq:2states}),  
$\mu_{RR} \,s_\xi^2$, and the observed values of the neutrino masses are achieved for  
$\mu_{RR}\sim 10$ keV and  $s^2_\xi \lesssim 10^{-6}$.

The only substantial difference from the main scenario
is the presence of three relatively light states $S_L$.
Therefore, in what follows we will focus on 
new physics associated to $S_L$.

According to (\ref{eq:addlagr}), the states $S_L$ have Yukawa interactions 
with the left leptonic doublet $L_L$ 
and heavy scalar doublet $\chi_L$ (\ref{eq:addlagr}). The mass of  $\chi_L$ is at
the $SU(2)_R$ symmetry breaking scale and  for estimations
we will use  $M_\chi = 3\cdot 10^5$ GeV. The Yukawa couplings $Y_L$
are large, being  $Y_{3} \sim 1$  for the third generation.
Furthermore, $S_{L}$ mix with light (mostly active) neutrinos according to  \cref{eq:2states}.
The mixing angles with light neutrinos  are given by
\begin{equation}
 \sin \theta_{Si} \approx  \frac{1}{\mu_{LLi}}(c_\xi \, m_{Di}' - s_\xi\, \mu_{LRi})\,, 
\label{eq:s-mix}
\end{equation}
where for simplicity we have taken the matrices $m_{D}'$,  $\mu_{LL}$ and $\mu_{LR}$ to be diagonal. 
In the case of full cancellation in the above expression, 
$S_L$ states and light neutrinos do not mix. 
The mixing gives additional contribution to the light neutrino masses 
\begin{equation}
\delta m_\nu \approx \sin^2 \theta_{Si}\, m_{LLi}\,. 
\label{eq:sscontr}
\end{equation}
Therefore, the condition that there is no significant contribution to the light neutrino masses from mixing with $S_L$ 
gives the upper bound  
\begin{equation}
\sin^2 \theta_{Si} \ll \frac{m_\nu}{m_{LLi}}\,, 
\label{eq:ineq}
\end{equation}
where ${m_\nu} \sim (0.01 - 0.02)$ eV.

The states $S_L$ are light sterile neutrinos and their properties (masses, mixing and interactions)  
are subject to strong cosmological bounds. At the same time, $S_L$ can be a dark matter candidate 
or even a very light particle in  meV -- eV mass range with no observable contributions to the energy density of the Universe.


The states $S_L$ decay into three light neutrinos, $S_{iL} \rightarrow \nu \nu \bar{\nu}$, 
via mixing with light neutrinos and $Z^0$ exchange. 
In vacuum, the lifetime equals
\begin{equation}   
\tau_i  = 3.8 \cdot 10^{20} \,{\rm sec} \,\left(\frac{10^{-6}}{\sin^2 \theta_{Si}}\right) 
\left(\frac{10~ {\rm keV}}{\mu_{LLi}}\right)^5. 
\label{eq:lifemix}
\end{equation}
Notice that, according to (\ref{eq:ineq}), for $\mu_{LLi} = 10$ keV and $\sin^2 \theta_{Si} = 10^{-6}$ 
the contribution to the light masses equals $\sim 0.01$ eV.  

The heavier $S_{iL}$ can also decay into lighter $S_{jL}$
with the $\chi^0$ exchange: $S_{jL} \rightarrow S_{iL} \nu \bar{\nu}$.
Typical time for this mode is 
\begin{equation}
\frac{1}{\Gamma_i} = 2.4  \cdot 10^{27} {\rm sec}  \left(\frac{10~ {\rm keV}}{\mu_{LLi}}\right)^5  
\left(\frac{M_{\chi}}{3 \cdot 10^5~ {\rm GeV}}\right)^4
\frac{1}{Y_j^2~ Y_i^2}\,,
\label{eq:lifeexc}
\end{equation}
which is much bigger than the decay time (\ref{eq:lifemix}) via process where mixing in employed.

Thus, the lifetime of $S_L$  with masses ($10$ - $100$) keV is much larger than the age of the Universe. 
Therefore, these $S_L$  can be candidates for the Dark matter particles  
if their appropriate number density is generated  
\cite{Dodelson:1993je,Shi:1998km,Merle:2013wta,Brdar:2017wgy,Asaka:2005pn,Asaka:2005an,Brdar:2015jwo,Baumholzer:2018sfb}.

$S_{iL}$ can be produced via mixing with active neutrinos
and oscillations. In this case the conditions on parameters of $S_{iL}$ to be a dark 
matter are similar to those in $\nu \text{MSM}$ \cite{Asaka:2005pn}. 
For masses in the ballpark of $10$ keV, 
the mixing should to be $\sin ^2 \theta_S  \lesssim  2.5\cdot 10^{-11}$. 
The strongest limits on $\theta_S$
arise from X-ray searches \cite{Perez:2016tcq,Abazajian:2001vt}, Supernova 1987A \cite{Raffelt:2011nc,Arguelles:2016uwb} 
and structure formation \cite{Schneider:2016uqi}. 

According to \cref{eq:s-mix}, for $m_D' = 0$ 
\begin{equation}
\sin \theta_S \approx -s_\xi \frac{\mu_{LR}}{\mu_{LL}}. 
\label{eq:s-mix1}
\end{equation}
Then,  for $s_\xi\simeq 10^{-3}$ \cref{eq:s-mix1} yields ${\mu_{LR}} \lesssim 10^{-2} \mu_{LL} $.

In addition, in our scenario $S_{iL}$ can be produced in the process
$\nu_i \bar{\nu}_j \rightarrow S_{kL} \bar{S}_{lL}$ with $\chi_L$-- exchange. 
Here, mixing is kept, so that $S_{kL}$ and ${S}_{lL}$ are the eigenstates of $\mu_{LL}$ matrix (with mixing due to $\mu_{LR}$ being neglected). 
Suppression of rate of this process, $\Gamma_S$, with respect to
the rate of usual active neutrino reactions, $\Gamma_\nu$, is 
\begin{equation}
\frac{\Gamma_S}{\Gamma_\nu} = \frac{4}{g^4}\, Y_{iL}^2 \, Y_{jL}^2 \,
|U_{ik}|^2|U_{jl}|^2 \, \left(\frac{m_W}{M_\chi}\right)^4\,
\approx  1.14\cdot 10^{-13}\,\, Y_{iL}^2  Y_{jL}^2\, |U_{ik}|^2 |U_{jl}|^2
\left(\frac{3 \cdot 10^{5} \,\text{GeV}}{M_\chi}\right)^4, 
\label{eq: production}
\end{equation}
where $U_{ik}$ is the mixing matrix elements of  $S_{kL}$ and $g$ is the weak coupling constant. 

Depending on values of masses and mixing of $S_{kL}$ one can consider different possibilities.  
Let us consider two extreme cases.\\

{\it{1.}} If the L-R symmetry holds in  the singlet sector, then  mixing between
$S_L$ states is the same as the one for $S_R$, which is of the TBM type. Since the mixing is large,  
the ratio of rates (\ref{eq: production}) will be determined by the largest coupling 
$Y_{3L} = 1$ and the mixing matrix elements $U_{3k} = U_{\tau k}^{\text{TBM}}$. 
If $S_{2L}$ is in the 10 keV range and other $S_L$ states are lighter, 
then $|U_{32}|^2 = 1/6$ and from (\ref{eq: production}) we obtain the ratio of rates 
${\Gamma_S}/{\Gamma_\nu} = 3 \cdot 10^{-15}$. Using this  ratio we find that $S_L$ exits the equilibrium with thermal bath 
at temperatures $T \sim 100$ GeV. Their density will be diluted  
due to decrease of number of degrees of freedom at lower temperatures. 
This dilution is, however, not enough and further suppression by factor of 
$30$ is needed to match correct  energy density of dark matter. 
This can be achieved if the reheating temperature, $T_\text{reh}$, after inflation is below  $100$ GeV.
Similar situation is for $S_{3L}$ and $S_{1L}$ which do not contribute substantially to the 
present energy density in the Universe due to smaller masses. 

With such low $T_\text{reh}$, however, it will be difficult to realize baryogenesis through 
leptogenesis, unless $M_\chi$ increases or the coupling $Y_L$ decreases leading to 
higher $S_L$ decoupling temperature. \\

{\it{2.}} If mixing in the $S_L$ sector is absent, the  DM component, 
taken to be  $S_{1L}$, will have the smallest coupling $Y_{1L} = 10^{-5}$. 
(This scenario can be reconciled with spontaneous L-R symmetry 
introducing additional flavons $\sigma$). 
In this case, the ratio of rates (\ref{eq: production}) equals
 $\approx 10^{-13} \,|Y_{1L}|^2 = 10^{-23}$. Correspondingly, the decoupling 
(freeze-out) temperature will be $\sim 10^5$ GeV and the required number density  
of $S_{1L}$ can be obtained at $T_\text{reh}\sim  10^{4}$ GeV. 
This is much higher than in the previous case, and opens
a possibility for low scale baryogenesis through leptogenesis.

Notice that in view of problems with generation of the
number density via mixing in $\nu\text{MSM}$ \cite{Asaka:2005pn}, the $\chi_L$-exchange
can be the main mechanism of the DM production, while the mixing
is suppressed sufficiently to satisfy the bounds from $X-$ray observations.





\section{Summary}
\label{sec:summary}
\noindent
The low scale left-right symmetric models accessible to the existing and planned colliders are
at odds with generation of  naturally small neutrino masses.
The ($1$ - $100$) TeV scale of the L-R symmetry breaking (and consequently, the scale of RH neutrino masses) requires very small  Dirac masses of neutrinos which strongly break the natural condition of quark-lepton similarity $Y \sim Y_q$. This similarity can be retained
using the inverse seesaw mechanism which requires introduction of three fermionic singlets
as well as the left and the right handed Higgs doublets.
These doublets break the L-R symmetry and provide the portal for interactions of singlets with the SM particles.

This setup allows to obtain the neutrino mixing of special form e.g. tribimaximal or bimaximal.
Under the screening condition, $Y_R = Y$, the light neutrino mass matrix is proportional to the Majorana mass matrix  $\mu$
of singlet fermions $S$. In turn,  special form of $\mu$ can be governed by symmetry in the S-sector. This symmetry is  generally broken by the
other interactions in the model but we have shown that the corrections
due to such breaking are small and do not destroy the structure of $\mu$.

 The contribution from the linear seesaw should be suppressed.
If dominant, it would require unnaturally small Dirac neutrino mass terms with the structure that breaks the $q$-$l$ similarity. 
Such suppression can be achieved by small VEV of the left-handed doublet.

The generic consequence of the inverse seesaw is the existence of
three heavy pseudo-Dirac neutral leptons with a mass splitting of the order of 
$\mu$. Under the conditions of  $q$-$l$ similarity and screening, these heavy leptons
mix with active neutrinos with  strength $\xi \simeq \langle\phi_1^0\rangle/\langle\chi_R^0\rangle$ 
and have strongly hierarchical mass spectrum. Consequently, only the lightest states, $N_1^\pm$, are accessible to current and near-future experiments. 
From the present experimental searches, we obtain the bound on mass of $N_1^\pm$ to be $M_1 \geq (0.4 - 2)$ GeV. This, in turn, 
leads to the lower bound on the scale of L-R symmetry breaking 
$\langle \chi_R^0\rangle > 200$ TeV. Future experiments such as SHiP, DUNE and FCC-ee can strengthen these bounds significantly. The FCC-hh collider will be able to test the existence of other new particles: $N_2^\pm$, $N_3^\pm$, the RH gauge bosons and new scalars.

Contributions of the heavy leptons to the effective Majorana mass $m_{ee}$  
are suppressed, being proportional to the contributions of light states. 

The leptogenesis scenarios do not yield the required amount of the baryon asymmetry 
in our model, primarily due to strong washout. However, with a certain extensions of the model (addition of new heavy fermion(s) $X$ and another singlet $S$ per generation) 
 or departure from $q$-$l$ similarity (ARS), the required lepton asymmetry can be produced. The electroweak baryogenesis is a viable option.
 
 The corrections to the Higgs mass ($\delta m^2_H$), induced via loops of RH neutrinos are, for $\langle \chi_R^0\rangle=200$ TeV, about 4 orders of magnitude larger than the experimentally observed value ($m_H^2\sim 10^4$ $\text{GeV}^2$).
The contributions from gauge boson loops as well as additional scalars
can lead to substantial reduction of this correction.
We derived the conditions for the complete cancellation of the Higgs mass one-loop corrections. Admittedly, such cancellations would be accidental.

Finally, we considered a scenario with two singlets per generation: $S_L$ in the left sector and 
$S_R$  in the right sector. In such a case the L-R symmetry is explicit and furthermore the global lepton number can be introduced.  
This modification leads to the appearance of  keV-scale leptons 
and the lightest of them can play the role of dark matter.

\section*{Acknowledgements}
\noindent
We are greatly indebted to Evgeny Akhmedov, Giorgio Arcadi, Arindam Das, Bhupal Dev, Peizhi Du, Oliver Fischer, Jisuke Kubo, Goran Senjanovi\'{c} and Xun-Jie Xu for several very useful discussions.

\bibliographystyle{JHEP}
\bibliography{refs}

\end{document}